\documentclass[fleqn,usenatbib]{mnras}

\usepackage[T1]{fontenc}
\usepackage{ae,aecompl}
\usepackage{courier}

\usepackage{graphicx}	
\usepackage{amsmath}	
\usepackage{amssymb}	
\usepackage{booktabs,array,tabularx}
\usepackage{epstopdf}

\makeatletter 
\def\eqalign#1{\null\,\vcenter{\openup\jot\m@th  \ialign{\strut\hfil$\displaystyle{##}$&$\displaystyle{{}##}$\hfil      \crcr#1\crcr}}\,} 
\makeatother

\DeclareMathAlphabet{\mathsc}{OT1}{cmr}{m}{sc}
\def\testbx{bx}%
\DeclareRobustCommand{\ion}[2]{%
\relax\ifmmode
\ifx\testbx\f@series
{\mathbf{#1\,\mathsc{#2}}}\else
{\mathrm{#1\,\mathsc{#2}}}\fi
\else\textup{#1\,{\mdseries\textsc{#2}}}%
\fi}

\title[Benchmark study of solar atomic models]{A benchmark study of atomic models for the transition region against quiet Sun observations}

\author[R.P. Dufresne et al.]{
R.P. Dufresne\thanks{E-mail: rpd21@cam.ac.uk},
G. Del Zanna,
and H.E. Mason.
\\
Department of Applied Mathematics and Theoretical Physics, University of Cambridge, Wilberforce Road, Cambridge CB3 0WA, UK\\
}

\date{Accepted XXX. Received YYY; in original form ZZZ}

\pubyear{2022}

\begin{document}
\label{firstpage}
\pagerange{\pageref{firstpage}--\pageref{lastpage}}
\maketitle

\begin{abstract}
The use of the coronal approximation to model line emission from the solar transition region has led to discrepancies with observations over many years, particularly for Li- and Na-like ions. Studies have shown that a number of atomic processes are required to improve the modelling for this region, including the effects of high densities, solar radiation and charge transfer on ion formation. Other non-equilibrium processes, such as time dependent ionisation and radiative transfer, are also expected to play a role. A set of models which include the three relevant atomic processes listed above in ionisation equilibrium has recently been built. These new results cover the main elements observed in the transition region. To assess the effectiveness of the results, the present work predicts spectral line intensities using differential emission measure modelling. Although limited in some respects, this differential emission measure modelling does give a good indication of the impact of the new atomic calculations. The results are compared to predictions of the coronal approximation and to observations of the average, quiet Sun from published literature. Significant improvements are seen for the line emission from Li- and Na-like ions, inter-combination lines and many other lines. From this study, an assessment is made of how far down into the solar atmosphere the coronal approximation can be applied, and the range over which the new atomic models are valid.
\end{abstract}




\section{Introduction}
\label{sec:intro}

Modelling line emission in the solar atmosphere becomes increasingly complex in the lower layers, as radiative transfer, (magneto-)hydrodynamics and various atomic processes become more important. (See Sect.\;\ref{sec:context} for more background on modelling line emission in the lower solar atmosphere.) Including all the relevant effects in self-consistent models makes the problem intractable. Inevitably, compromises are made in some areas to favour exploration in another: such as, radiative transfer calculations have been made in hydrostatic equilibrium \citep[e.g.][]{skelton1982, lanzafame1994}, while investigation of new atomic models have been carried out in optically thin conditions and/or steady state equilibrium \citep[e.g.][]{baliunas1980, nussbaumer1975}. More recently, emphasis has been placed on exploring radiative transfer and dynamical effects together, but there have been compromises in the atomic models. This includes, for example, treating H and He in local thermodynamic equilibrium \citep[such as][]{martinez2016} or using simplified atomic models and approximations for atomic cross sections \citep[like][who use six levels for H and three for He]{golding2016}.

While more complex atomic models have been explored in isolated cases for the solar transition region, atomic modelling designed for the corona is most commonly used by theoreticians and those interpreting spectroscopic observations. The assumptions used in coronal atomic models result in ion fractions which are independent of density. This allows the use of just one table of ion fractions when modelling a variety of situations. The result has been a number of these tables being produced over the years \citep[e.g.][]{arnaud1985, mazzotta1998}. Such atomic data have been integrated into radiative transfer and hydrodynamic models, as well as into large scale, synthetic spectral codes.

Even in the interpretation of some of the first ultra-violet (UV) observations from rocket flights, however, \citet{burton1971} found, when they used coronal atomic models, that emission from Li- and Na-like ions in the transition region differed notably from the emission measure of most other ion sequences. A similar problem also occurs in stellar atmospheres \citep{delzanna2002}. For Li- and Na-like ions, the primary focus in atomic models has been to add the effects of density on atomic processes included in the coronal approximation \citep[e.g.][]{vernazza1979, doyle2005}. In some cases, tables of ion fractions which include density effects have been made available for this purpose, such as those by \citet{jordan1969} and \citet{summers1974}. However, \citet{judge1995} included this effect and believed it insufficient to account for the discrepancies.

\citet{doschek1999} investigated other low charge ions in the transition region (TR) using coronal atomic models in an isothermal approximation. They found differences by factors of two to five between predictions and observations for \ion{C}{ii}, \ion{N}{iii} and \ion{S}{iii-v}, although the observations were limited as a result of the lines not being observed simultaneously. Their conclusion was that inaccurate excitation data at the time, much of which did not include resonant excitation, was the cause of the discrepancies.

The transition region clearly requires an atomic modelling framework that is distinctive from that of the corona. The aim of the present work is to benchmark new atomic models which have been designed for the TR. The models were built over a series of four works \citep{dufresne2019, dufresne2020, dufresne2021pico, dufresne2021picrm}, and assessed the effect on ion formation and level populations of atomic processes not normally included in the coronal approximation. The complete models synthesise processes which have been shown to affect transition region ions: they include density suppressing dielectronic recombination \citep{burgess1969}; density enhancing electron impact ionisation \citep[][for carbon]{nussbaumer1975}; the solar radiation causing photo-ionisation \citep[also][]{nussbaumer1975}; as well as including collisions with hydrogen and helium \citep[][for silicon]{baliunas1980}. The new atomic models highlight that all of these processes have an effect on the more abundant elements observed in the transition region (C, N, O, Ne, Mg, Si, S). The benchmarking in this work is carried out by comparing predicted intensities from the models to predictions from coronal models and to observations.

A comprehensive model of transition region line emission is clearly a vast undertaking and beyond the scope of this paper. In their conclusions, \citet{judge1995} listed a raft of other factors which may by required to fully account for emission from the transition region: blends and unidentified lines, ion diffusion, global and local elemental abundance variations, instrumental calibration, optical depth, time-dependent ionisation, hydrogen absorption and non-Maxwellian electrons. Despite this, the present study will help inform which atomic processes are required for more complete models of line emission in the TR. The ion fractions from the present atomic models have been made available for such purposes.

The next section gives a brief description of the atomic models, the methods used to predict line intensities, and the sources of the solar observations used for comparison. Section\;\ref{sec:results} gives intensities predicted by the models for key lines and compares them with predictions from the coronal approximation and with observations. The conclusions in Sect\;\ref{sec:concl} summarise the findings and give an indication at which point in the atmosphere the present models should be used in preference to coronal models. In the Appendix, predicted intensities for a number of additional lines are given, as well as an assessment of how much varying the model parameters could affect the results.

\section{Methods}
\label{sec:methods}

\subsection{Background on modelling line emission}
\label{sec:context}

Whether more advanced modelling than the coronal approximation is required for line emission in the solar atmosphere depends very much on the lines and conditions being investigated. \citet{hansteen1993} included time-dependent ionisation and recombination while modelling heating and cooling due to a succession of nanoflares. It was found that emission from ions with long ionisation and recombination times are affected the most because upper energy levels are enhanced through excitation before transitions to another charge state takes place. This particularly affects Li-like ions, such as \ion{O}{vi} and \ion{C}{iv}, while emission from ions with short ionisation and recombination times, such as \ion{O}{iv}, is similar to the assumption of ionisation equilibrium. \ion{C}{iv} ion fractions were altered by a factor of three during heating and cooling phases compared to the ionisation equilibrium in coronal models. The total radiative losses in parts of the TR were a factor of two higher.

\citet{pietarila2004} investigated the effects of ion diffusion on emission. They used similar atomic models as the present work, and measured the enhancements to line emission relative to ionisation equilibrium. They found the effects of time dependent ionisation are stronger for transitions involving $\Delta{\rm n}\geq 1$, where $n$ is the principal quantum number. Enhancements of 2-4 in line emission occurred in quiet Sun conditions for many lines from C, O and Si, such as the \ion{C}{ii} 687.05\,\AA\ and \ion{Si}{iv} 818.15\,\AA\ lines. For He lines, which all involve $\Delta{\rm n}\geq 1$ transitions, the enhancements were 7-10. The enhancement was typically 1.5-2 for $\Delta{\rm n}=0$ transitions, which are the majority of the lines observed in the TR. Despite these enhancements, \citet{pietarila2004} caution about the lack of evidence seen in  observations for ion diffusion, and the number of questionable assumptions involved in modelling it.

\citet{pietarila2004} also carried out one-dimensional, radiative transfer calculations, which model self-consistently the number of escaping photons and centre-to-limb variation. They assumed hydrostatic conditions, and took model atmosphere parameters and hydrogen fractions for the lower atmosphere from \citet{vernazza1981}; coronal radiation was used for the upper boundary. Although much of the emission in the TR is optically thin, they found disc-centre, emergent intensities could be enhanced by up to a factor of three due to photon scattering for some lines with low optical depth, such as \ion{C}{ii} 1334.53\,\AA\ and \ion{Si}{iii} 1206.51\,\AA\ lines. In their models, He lines, which show the biggest discrepancies in the TR between observations and coronal models, were shown to have very high optical depths. This produced enhanced emission at disc centre due to photon scattering in the radial direction, and neither brightening nor darkening at the limb in the He lines. Compared to their optically thin calculation, the emergent intensities of the He resonance lines were enhanced by up to a factor of four.

There are still other factors taking place in the solar atmosphere which could help resolve the discrepancies in line emission. Non-Maxwellian electron distributions can alter diagnostics in two ways. They can change the temperatures and densities at which ions form. This, in turn, alters the interpretation of line ratios, \citep[see as an example][]{dudik2014}. They are also likely to enhance lines emitted from levels with an high excitation energy, such as the $\Delta{\rm n}\geq 1$ transitions relevant for time-dependent ionisation, \citep[see e.g.][]{dzifcakova2011}. Extending radiative transfer and hydrodynamic calculations to two and three dimensions are also contributory factors to resolving discrepancies, \citep[such as][]{rathore2015}. The main factors noted in the previous paragraphs, however, highlight the primary factors to consider. Of equal importance to these factors is the need for high quality atomic models; this is the focus of the rest of this work.

\subsection{Atomic models}

The assumptions used for the coronal approximation apply to conditions typically prevalent in the solar corona: the plasma is optically thin; collisions with ions are weak so that only collisions with electrons need be considered, plus radiative decay (and, in some cases, proton collisions affecting fine-structure levels); plasma time-scales allow ionisation equilibrium to be applied; the upper energy levels of the majority of lines which produce coronal emission are populated through collisions with electrons instead of recombination; effective ionisation and recombination rate coefficients do not change with density. All of the assumptions make it possible to calculate ion formation separately from energy level populations. Ion fractions are also independent of density in the approximation, allowing tables of ion fractions to be pre-calculated.

The ion balances of \citet{shull1982} and \cite{arnaud1985}, for example, used approximations for ionisation and recombination rates and have been used for many years. When large scale, \textit{ab initio} calculations for ionisation and recombination rates became available, they were finally supplanted by new tables \citep[such as][]{bryans2006, dere2009}. The latest version of the \textsc{Chianti} database \citep[v.10][]{dere1997, delzanna2021v10} allows the user to create a table of ion fractions that includes an estimate of how effective recombination rates are suppressed with density \citep[using the work of][]{nikolic2013, nikolic2018}, but the default remains the density-independent ion balances.

In this work, the default ion balances of \textsc{Chianti} v.10 are used to compare with the new atomic models for the transition region. The latter models are described in detail in \citet{dufresne2021pico} for C and O, and in \citet{dufresne2021picrm} for N, Ne, Mg, Si and S. In short, the models include: the change in effective ionisation rates as metastable levels become populated at higher density; an estimate of suppression of effective recombination rates with higher densities using the tables of \citet{summers1974}; photo-ionisation using a fixed radiation field derived from the irradiances of \citet{woods2009}; and, charge transfer with hydrogen and helium. There are two model types for each element. The first are referred to as \textit{electron collisional models (e)}; they include only the effects of higher density on electron impact ionisation and radiative and dielectronic recombination. The second set of models are termed \textit{full models (f)}, and include the effects of photo-ionisation (PI) and charge transfer (CT), as well as the effects included in the electron collisional models. The models were run at a constant pressure of 3$\times$10$^{14}$\,cm$^{-3}$\,K, which is consistent with the model atmosphere of \citet{avrett2008}, from which the hydrogen abundances were taken to calculate charge transfer rates.

\subsection{Predicting line intensities}
\label{sec:demmethods}

In a multi-thermal atmosphere with an electron temperature $T$, electron number density $N_e$ and assuming that emission is optically thin, the intensity along the line of sight of a line with wavelength $\lambda_{ji}$, emitted from upper state $j$ to lower state $i$, is given by

\begin{equation}
	I_{ji}~=~Ab(Z)\int_T ~C(T,\lambda_{ji},N_e)\; N_e\, N_H ~dh\; ,
	\label{eqn:intens}
\end{equation}

\noindent where $Ab(Z)$ is the abundance of element Z relative to hydrogen, $N_H$ is the hydrogen number density, $h$ is height through the atmosphere, and

\begin{equation}
	C(T,\lambda_{ji},N_e)~=~\frac{h\nu}{4\pi N_e}~A_{\lambda_{ji}}N^{+q}_{Z,j}
	\label{eqn:contrib}
\end{equation}

\noindent is the contribution function of the line, where $A_{\lambda_{ji}}$ is the spontaneous decay rate and $N^{+q}_{Z,j}$ is the number density of the relevant ion in the upper level from which the line is emitted. (The separate treatment of level populations and ion fractions in the coronal approximation allows the substitution $N^{+q}_{Z,j}=\frac{N^{+q}_j}{N^{+q}}\frac{N^{+q}}{N_z}N_Z$.) The electron temperature is assumed to be the same as the ion temperature in the present models.

The spatially inhomogeneous and dynamic nature of the Sun means it is difficult to determine electron and hydrogen densities along the line of sight. Such quantities have been determined in some works through one-dimensional, hydrostatic radiative transfer calculations \citep[such as][]{avrett2008, fontenla2014}, where certain parameters have been imposed on the models to bring better agreement with observations. The calculations have shown that there is a correlation between height and temperature, so that it is possible to define a new quantity, $DEM(T)$, the differential emission measure. The integration may now be carried out over temperature. Line intensity is then given by

\begin{equation}
	I_{ji}~=~Ab(Z)\int_T ~C(T,\lambda_{ji},N_e)\; DEM(T)\;dT\; .
	\label{eqn:dem}
\end{equation}

Many authors have sought to determine the differential emission measure (DEM) by using observations of lines formed at various temperatures along the line of sight. This is carried out by inverting equation\;(\ref{eqn:dem}) for each line by using its contribution function and observed intensity. Many inversion routines exist, but the feasibility of the inversion has been in dispute for a long time because the solutions are not unique, it assumes a single-valued temperature at each height and that the atmosphere is static, \citep[as discussed in such works as][]{craig1976, judge1997}. More practical solutions with constraints have been developed in light of these issues, and many authors have shown that, broadly, the observed intensities can be well represented, with some notable exceptions. \citet{delzanna2018} give an overview of a number of the issues relating to this topic.

The resulting DEM will inevitably depend on the lines used and the atomic data fed in. In this work, it can be used as a tool to assess the different atomic models by keeping everything else constant in the calculation and only changing the ionisation equilibria. The atomic model that is a more suitable description of solar conditions will be the one that has predicted to observed intensities in closest agreement within an ion and with other lines formed at a similar temperature. The DEM routine, \texttt{chianti\_dem}, from \textsc{Chianti} v.10 was used to predict intensities, and the photospheric abundances of \citet{asplund2009} were used throughout. Both the transition region models and the coronal approximation in this work use the same excitation and radiative decay data, which means that it is the ion balances that are being tested. Also given in the results is the effective temperature of a line, defined by

\begin{equation}
	T_{\rm eff} \,=\, \frac{\int C(T,\lambda_{ji},N_e)\, DEM(T) \,T\;dT} {\int C(T,\lambda_{ji},N_e) \; DEM(T) \;dT} \;.
	\label{eqn:efftemp}
\end{equation}

\noindent This is an average temperature more indicative of where a line is formed, but is only valid in certain cases, such as in plane parallel atmospheres.

\subsection{Observational data}
\label{sec:obsmethods}

In this paper, theoretical intensities from the new atomic results are benchmarked against average, quiet Sun observations. To fit the DEM and test lines over the whole transition regions requires observations from lines forming over a broad wavelength and temperature range. This requires the use of data obtained from different instruments on different dates published by different authors. In several instances, even when lines were observed by the same instrument, they were not recorded simultaneously. This, in addition to the large variability of the quiet Sun in TR lines, means that in many instances there is a large scatter of averaged radiances in the literature, as detailed below. Nevertheless, by using the same set of observed radiances but different atomic models, it is possible to show the way in which these more advanced models alter predictions.

The source for the majority of the lines comes from \citet{warren2005}, who used the Solar Ultraviolet Measurements of Emitted Radiation (SUMER) instrument and the Coronal Diagnostic Spectrometer (CDS), both on the Solar and Heliospheric Observatory (SOHO). All lines from that work below 650\,\AA\ are from CDS and those above are from SUMER. Observations are required for longer wavelength lines. Line radiances were measured from the atlas of \citet{brekke1993}, who used the High Resolution Telescope and Spectrograph (HRTS), which covered the range 1190-1730\,\AA. \citeauthor{brekke1993} produced data from a variety of regions, including four of the quiet Sun. The intensities in this work are taken from the region labelled QR\;A; this area is far from the active region present at the time. The values from this region are mostly within 40 per cent of those recorded by \citeauthor{warren2005}, for the lines they have in common.

A few additional observations were taken from: \citet{parenti2005}, \citet{wilhelm1998} and \citet{pinfield1999} (all using SUMER); \citet{vernazza1978} and \citet{nicolas1977} (both from Skylab); and, \citet{andretta2014}, who used CDS. Measurements carried out within the same year by the same instrument \citep{warren2005, wilhelm1998, parenti2005} typically agree to within 40 per cent. Lines from \citet{sandlin1986} were also checked, but in many cases greater discrepancies were found compared to the other works listed above and they were not considered further. 

The majority of the intensities for the lines chosen here are from quiet Sun conditions close to solar minimum. The observations represent an average intensity, but it should be remembered that: fluctuations are seen in emission over time; images of transition region lines show the influence of active regions and granulation patterns; and, emission varies between cell centres and network lanes. Preference was given to authors whose measurements showed the broadest agreement overall for lines within the same ion. Next, preference was given to those who reported several lines within a multiplet. This helps gain an idea of whether individual lines may be affected by blends and whether opacity effects are present.

\subsection{The derived differential emission measure}

\begin{figure}
	\centering
	\includegraphics[width=8.4cm]{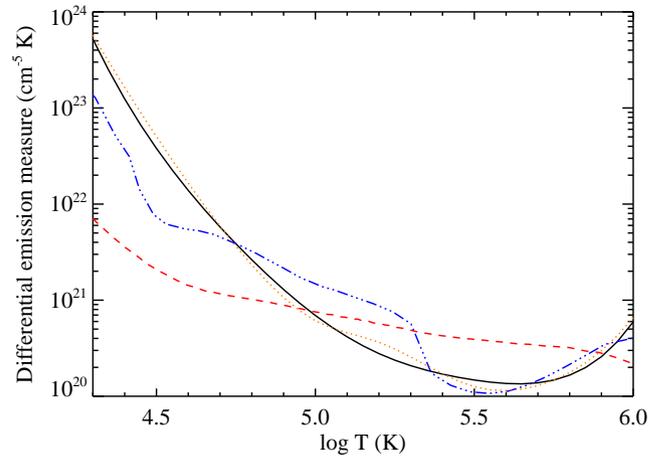}
	\caption[DEMs derived from various models]{Comparison of the DEMs derived from the present work with those from model atmospheres; black solid line - full models, orange  dotted - coronal approximation, red dashed - DEM from \citeauthor{avrett2008}, blue dot-dot-dashed - DEM from \citeauthor{fontenla2014}.}
	\label{fig:dem}
\end{figure}

Relatively strong lines with few blends were used to fit the DEM. No Li- or Na-like lines were used (apart from \ion{Ne}{viii}), nor those which were far from the main trend of other lines. Only elements with high first ionisation potential (FIP) were used to fit the DEM in the transition region, to avoid complications due to any potential FIP bias. To fully calculate emission from lines forming high in the transition region requires calculating the DEM at coronal temperatures. Only observations from low FIP elements are generally available at these temperatures, and so lines from low FIP elements were used to constrain the DEM at temperatures above 1\,300\,000\,K. However, these lines are not included further in the discussion. Similarly, lines from atoms were included at low temperatures (<\,10\,000\,K) purely to constrain the DEM for low charge ions. Other than that, neutral lines are not considered further.

The observed lines used to fit the DEM for each of the models are marked by an asterisk in the tables of results throughout the main paper and appendix. The extra lines used to constrain the DEM at the lowest and highest temperatures are shown in Tab.\;\ref{tab:demlines}. The resulting DEMs from the different models are very similar across the whole transition region (${\rm log}\,T \approx 4.4-5.8$), even though the ion fractions have shifted to lower temperature with the improved atomic models. The DEMs derived from the full model and the coronal approximation are shown in Fig.\;\ref{fig:dem}. The main differences in the DEMs for the various models begin to show only in the upper chromosphere, where the greatest changes in the ion balances occur. The DEMs are also compared with those derived from the model atmospheres of \citet{avrett2008} and \cite{fontenla2014}. The latter is from their model B for cell centre, and there is a jump at ${\rm log}\,T=5.3$ because their separate TR and coronal models have been joined together for this illustration. While there are differences in detail, the present DEMs broadly follow the DEM from \citet{fontenla2014}.

\begin{table*}
	\caption{Comparison of predicted and observed quiet Sun radiances for ions forming at the transition region-corona boundary.}
	\centering
	\begin{tabular}{p{1.3cm}>{\raggedright}p{0.7cm}>{\raggedleft}p{1.3cm}>{\raggedleft}p{1.3cm}>{\centering}p{0.7cm}>{\raggedright}p{0.8cm}>{\centering}p{0.7cm}>{\centering}p{0.7cm}>{\centering}p{0.7cm}>{\centering\arraybackslash}p{1.9cm}}
			\hline\hline \noalign{\smallskip}
			Ion & Seq & $\lambda_{\rm obs}$ & $I_{\rm obs}$ & $T_{\rm c}$ & $T_{\rm f}$ & $R_{\rm c}$ & $R_{\rm e}$ & $R_{\rm f}$ & $R_{\rm f}^{\rm alt}$ \\
			\noalign {\smallskip} \hline \noalign {\smallskip}

\ion{Mg}{vii} & C & $^{bl}$435.20 & $28.3^f$ & 5.81 & 5.78 & 0.68 & 0.59 & 0.60 &  \\
\ion{Mg}{vii} & C & $^{sb}$365.20 & $10.0^a$ & 5.90 & 5.88 & 0.73 & 0.60 & 0.63 &  \\
\ion{Mg}{vii} & C & $^{sb}$367.67 & $21.0^a$ & 5.90 & 5.87 & 0.58 & 0.48 & 0.49 &  \\
\ion{Mg}{viii} & B & $^{sb}$436.70 & $42.7^f$ & 5.98 & 5.97 & 0.72 & 0.57 & 0.59 &  \\
\ion{Mg}{viii} & B & 313.77 & $30.1^a$ & 5.99 & 5.99 & 0.61 & 0.47 & 0.49 &  \\
\ion{Mg}{viii} & B & 315.02 & $71.8^a$ & 5.99 & 5.98 & 0.60 & 0.47 & 0.49 &  \\
\ion{Mg}{viii} & B & 338.99 & $14.9^a$ & 5.99 & 5.98 & 0.69 & 0.54 & 0.57 &  \\
\ion{Mg}{viii} & B & $^{bl}$317.01 & $20.5^a$ & 6.00 & 6.00 & 0.70 & 0.57 & 0.59 &  \\
\ion{Mg}{viii} & B & $^{bl}$335.31 & $24.8^a$ & 6.07 & 6.06 & 0.62 & 0.51 & 0.52 &  \\
\ion{Si}{viii} & N & 314.37 & $27.3^a$ & 6.03 & 6.03 & 0.68 & 0.60 & 0.61 &  \\
\ion{Si}{viii} & N & 316.22 & $45.9^a$ & 6.03 & 6.03 & 0.80 & 0.70 & 0.72 &  \\
\ion{Si}{viii} & N & 319.83 & $69.2^a$ & 6.03 & 6.03 & 0.79 & 0.70 & 0.71 &  \\
			\noalign{\smallskip}\hline
			\noalign{\smallskip}
    \multicolumn{10}{p{0.8\linewidth}}{\textbf{Notes.} Ion - principal emitting ion, and `*' denotes a line used to fit the DEM; Seq - ion isoelectronic sequence; observed wavelength $\lambda_{\rm obs}$ (\AA), where superscript `sb' denotes a self-blend and `bl' a blend; the measured radiance $I_{\rm obs}$ (ergs cm$^{-2}$ s$^{-1}$ sr$^{-1}$) using: a) \citeauthor{warren2005}, b) \citeauthor{brekke1993}, c) \citeauthor{wilhelm1998}, d) \citeauthor{parenti2005}, e) \citeauthor{pinfield1999}, f) \citeauthor{vernazza1978}, g) \citeauthor{nicolas1977}, and h) \citeauthor{andretta2014}; $T$ - the effective temperature for each line (logarithmic values, in K); $R$ - the ratio between the predicted and observed intensities; subscripts of $T$ and $R$ refer to results obtained using: c) \textsc{Chianti} coronal approximation ion fractions, e) ion fractions from electron collisional models, and f) ion fractions from full models; and, $R^{\rm alt}_{\rm f}$ - the range of $R_{\rm f}$ using the highest and lowest observations from the other sources listed.}
		\end{tabular}
	\normalsize
	\label{tab:trcorona}
\end{table*}

\section{Results}
\label{sec:results}

Each predicted line intensity is expressed as a ratio, $R$, with the observed intensity of the line, where $R_{\rm c}$ are those from the coronal approximation of \textsc{Chianti}, $R_{\rm e}$ are from the electron collisional models, and $R_{\rm f}$ are from the full models. There are relatively few observations of long wavelength, inter-combination lines from low charge ions. In these cases, estimated intensities were obtained by using their observed intensity ratios with known lines and multiplying it by the disc centre intensity of the reference line. All of the intensity ratios were observed close to the limb, and different lines have different centre-to-limb enhancements. To reduce the uncertainty, reference lines were chosen which had similar limb brightening as the unknown line. Testing this approach on lines with known intensities gave estimates within 1.5-2.0 of actual values.

If the predicted line intensity has contributions from other ions of more than 10 per cent it is indicated as a blend, or as a self-blend if such contributions come from the same ion. The effective temperatures are given for the coronal approximation, $T_{\rm c}$, and the full model, $T_{\rm f}$. Comparison of these two highlights the largest change in effective temperature exhibited by the models. $R^{\rm alt}_{\rm f}$ labels what the lowest and highest values of $R_{\rm f}$ would be if observed intensities from the other sources given in Sect.\;\ref{sec:obsmethods} were used. This gives an indication of the level of uncertainty present in the observations. Isoelectronic sequence for each ion is also given, allowing an assessment of whether there are trends in the results for any of the sequences.

\subsection{The transition region-corona boundary}
\label{sec:trcorona}

In \citet{dufresne2021picrm} it was shown that the formation of the higher charge states of Mg, Si and S (Ne-like and above) are barely altered by the new effects added to the models. This is seen in the predicted intensities here, which change by 10 per cent for Si and 10-20 per cent for Mg, as shown in Tab.\;\ref{tab:trcorona}. The intensities all decrease because ion formation moves to lower temperature and the emission measure is downward sloping in that direction. The coronal approximation and the present atomic models for the TR give much the same description of emission, as expected.

There is good consistency in the predicted to observed ratios for both \ion{Mg}{vii} and \ion{Mg}{viii}. In this region, the only observations from high-FIP elements available to fit the DEM are from Ne. (See Tab.\;\ref{tab:appuppertr} for the results for the Ne lines). Since the ratios for the Mg lines lie approximately in the range 0.5-0.6, it suggests a possible FIP bias of about 1.8 for Mg relative to Ne compared to the photospheric abundances of \citet{asplund2009}. The bias is close to the value of 1.6 obtained by \citet{young2018}; \citeauthor{young2018} used the same pressure, the coronal approximation of \textsc{Chianti} and an estimate of DR suppression with density. \citeauthor{young2018} only assessed Mg lines, but assumed the bias would be similar for Si. It is noted that the ratios here for \ion{Si}{viii} lie in the range 0.6-0.7, which suggest a possible FIP bias of approximately 1.5 for Si. This bias is consistent with the first ionisation potential of Si being higher than that of Mg.

\subsection{The upper transition region}
\label{sec:uppertr}

For this work, the upper TR is broadly defined to be above 100\,000\,K; it typically includes lines from ions with charge $+3$ and higher. (Lines from Li- and Na-like ions will be treated separately in the next section.) The only changes to ion formation in this region shown by the present atomic models, compared to the coronal approximation, are due to density suppressing DR and enhancing ionisation by electron impact. All of the ions in this region form at lower temperature and some increase in peak abundance relative to the coronal approximation. This can have a significant impact on diagnostics, such as when using line ratios. This has been demonstrated recently by \citet{rao2022pl}, who illustrated, using the same ion balances as the present work, how the \ion{O}{iv} 1401.15\,\AA\ / \ion{S}{iv} 1406.04\,\AA\ line ratio changes by factors of two to four in isothermal conditions.

Because the DEM in this region is relatively flat, the integrated line intensities do not change as much as line ratios, despite the shift in ion formation to lower temperatures. The lines intensities most affected are those of inter-combination lines emitted from levels close in energy to the ground. This is because the lines form at lower temperature than dipole-allowed lines emitted from the same ion. The inter-combination line intensities in this region are enhanced by up to 35 per cent, as shown in Tab.\;\ref{tab:uppertr}.

Using the ion fractions from the present modelling to fit the DEM means that the predicted intensities for the \ion{O}{iv} inter-combination lines are slightly closer to the emission measure of other ions than in the results of \citet{dufresne2020}. In that earlier work, coronal ion fractions were mostly used to fit the DEM. There are still differences in the ratios for each of the inter-combination lines from this ion. Since the strongest line within the multiplet is under-predicted the most, it suggests that redistribution of intensity to other lines within the multiplet is not the cause of this. Perhaps it is the time-dependent ionisation effects that \citet{olluri2013} investigated that account for the different ratios within this density sensitive multiplet. In their models, the lines form over a much wider height and temperature range than predicted when assuming ionisation equilibrium.

Such non-equilibrium ionisation effects may also account for the discrepancy that is still present in the \ion{O}{v} inter-combination line. Another factor could be due to it being weak and very close in wavelength to the hydrogen Lyman-$\alpha$ line, making it difficult to separate from the Lyman-$\alpha$ wings.

\begin{table*}
	\caption{Comparison of predicted and observed quiet Sun radiances for inter-combination lines forming in the upper TR.}
	\centering
	\begin{tabular}{p{1.3cm}>{\raggedright}p{0.7cm}>{\raggedleft}p{1.3cm}>{\raggedleft}p{1.3cm}>{\centering}p{0.7cm}>{\raggedright}p{0.8cm}>{\centering}p{0.7cm}>{\centering}p{0.7cm}>{\centering}p{0.7cm}>{\centering\arraybackslash}p{1.9cm}}
			\hline\hline \noalign{\smallskip}
			Ion & Seq & $\lambda_{\rm obs}$ & $I_{\rm obs}$ & $T_{\rm c}$ & $T_{\rm f}$ & $R_{\rm c}$ & $R_{\rm e}$ & $R_{\rm f}$ & $R_{\rm f}^{\rm alt}$ \\
			\noalign {\smallskip} \hline \noalign {\smallskip}

\ion{S}{iv} & Al & 1406.04 & $3.5^b$ & 5.00 & 4.89 & 0.61 & 0.77 & 0.82 &  \\
\ion{N}{iv}* & Be & 1486.54 & $13.2^b$ & 5.13 & 5.04 & 0.77 & 0.98 & 0.95 &  \\
\ion{O}{iv} & B & 1399.76 & $5.1^b$ & 5.17 & 5.08 & 0.70 & 0.83 & 0.84 & 0.71 \\
\ion{O}{iv} & B & 1401.15 & $32.3^b$ & 5.18 & 5.09 & 0.62 & 0.72 & 0.72 &   0.64 \\
\ion{O}{iv} & B & 1404.77 & $13.0^b$ & 5.18 & 5.09 & 0.80 & 0.90 & 0.90 &    \\
\ion{O}{iv} & B & 1407.36 & $3.1^b$ & 5.18 & 5.08 & 1.12 & 1.32 & 1.33 &  \\
\ion{S}{v} & Mg & 1199.18 & $6.5^d$ & 5.19 & 5.11 & 0.55 & 0.70 & 0.68 &  \\
\ion{O}{v} & Be & 1218.34 & $89.7^a$ & 5.35 & 5.32 & 0.40 & 0.38 & 0.40 & 0.23 - 0.57 \\
			\noalign{\smallskip}\hline
			\noalign{\smallskip}
    \multicolumn{10}{p{0.8\linewidth}}{\textbf{Notes.} Same as Table\;\ref{tab:trcorona}.}
		\end{tabular}
	\normalsize
	\label{tab:uppertr}
\end{table*}

The \ion{S}{v} inter-combination line at 1199.18\,\AA\ appears to be further from observations than the others, but the SUMER observation includes a second order blend from \ion{O}{iii}, which has not been included in the predicted intensity. The \ion{S}{v} lines are those for which \citet{doschek1999} found the greatest discrepancy with observations, up to a factor of five. This is not the case here, and the reasons are discussed further in Sect.\;\ref{sec:doschek} of the Appendix. All of the predicted intensities for the dipole allowed lines forming in this region are altered by 15 per cent at the most; the results for those lines are given in Sect.\;\ref{sec:appuppertr} and Tab.\;\ref{tab:appuppertr}.

\subsection{Li- and Na-like ions}
\label{sec:lilike}

As highlighted in Sect.\;\ref{sec:intro}, the discrepancies in predictions from the coronal approximation for all Li- and Na-like ions forming in the TR have been noted for a long time. Such discrepancies exist regardless of the temperature at which they form. The improvements in predictions for these ions with the present modelling are some of the largest seen throughout this work. Apart from \ion{Si}{iv}, the changes are all a result of density effects on DR and electron impact ionisation.

For the resonance lines of \ion{C}{iv}, observations from \citet{brekke1993} were used because they are available for both the 1548.20\,\AA\ and 1550.82\,\AA\ lines. The factor of three enhancement in these lines is the same as the results shown in \citet{dufresne2019}. The intensities used for \ion{C}{iv} come from the region labelled QR by \citet{brekke1993}, instead of the QR A region used for all other observations. The intensities of both these lines are four times stronger in the QR A region. The intensity given in \citet{wilhelm1998} for the 1548.20\,\AA\ line lies halfway between the two \citeauthor{brekke1993} observations, and provides the alternative ratio of 0.74 shown in Tab.\;\ref{tab:lilike}.

Despite the improved predictions for the Li-like \ion{O}{vi} lines, their emission measure are still a factor of two lower than for other lines used to determine the DEM. The alternative observations of \citet{wilhelm1998} and \citet{parenti2005} are closer to the current predictions, but the values from \citet{vernazza1978} are closer to \citeauthor{warren2005}. The increase in the predicted intensities for the \ion{N}{v} lines is the same as that estimated by \citet{doyle2005} when including density effects in the models. The changes, however, are insufficient to account for the high observed intensities. There is a factor of two difference between the alternative intensities of \citeauthor{brekke1993} and \citeauthor{wilhelm1998} for these lines, but the predicted intensities are still far short of all the observations.

The Na-like \ion{S}{vi} line at 944.55\,\AA\ has a blend with \ion{Si}{viii}, explaining the higher effective temperature and why the prediction using the present modelling does not change as much as the other resonance line. By estimating what the contribution from \ion{Si}{viii} would be, based on the results seen above in Sect.\;\ref{sec:trcorona}, and subtracting these from the predictions and observations, gives $R_{\rm f}$ of the \ion{S}{vi} line to be approximately 0.5. This is still quite far from the other \ion{S}{vi} resonance line at 933.40\,\AA, which may indicate there is a blend in the latter line. The lines show similar discrepancies as the Li-like \ion{N}{v} and \ion{O}{vi} lines and the Na-like \ion{Si}{iv} lines when using the observations of \citet{warren2005}. The other two \ion{S}{vi} lines are extremely weak and are included purely to verify the results for the resonance lines; they show broad agreement in the ratios for this ion.

Emission from Na-like \ion{Si}{iv} has been discussed for many years, including the early analysis of \citet{burton1971} mentioned in Sect.\;\ref{sec:intro}. There is evidence in the QS that the ratio of the doublet is close to two and the lines are optically thin, \citep[see as examples][]{kerr2019, rao2022ntv}. Here, it can be seen that the full models improve the predictions for these lines by almost a factor of six. This is due to charge transfer with He and density effects on electron collisional processes substantially enhancing the fractional abundance of this ion.

\begin{table*}
	\caption{Comparison of predicted and observed quiet Sun radiances for Li- and Na-like ions.}
	\centering
	\begin{tabular}{p{1.3cm}>{\raggedright}p{0.7cm}>{\raggedleft}p{1.3cm}>{\raggedleft}p{1.3cm}>{\centering}p{0.7cm}>{\raggedright}p{0.8cm}>{\centering}p{0.7cm}>{\centering}p{0.7cm}>{\centering}p{0.7cm}>{\centering\arraybackslash}p{1.9cm}}
			\hline\hline \noalign{\smallskip}
			Ion & Seq & $\lambda_{\rm obs}$ & $I_{\rm obs}$ & $T_{\rm c}$ & $T_{\rm f}$ & $R_{\rm c}$ & $R_{\rm e}$ & $R_{\rm f}$ & $R_{\rm f}^{\rm alt}$ \\
			\noalign {\smallskip} \hline \noalign {\smallskip}

\ion{Si}{iv} & Na & 1393.78 & $280.0^a$ & 4.90 & 4.76 & 0.06 & 0.17 & 0.35 & 0.27 \\
\ion{Si}{iv} & Na & 1402.77 & $127.0^a$ & 4.90 & 4.76 & 0.07 & 0.19 & 0.39 & 0.30 \\
\ion{Si}{iv} & Na & 1128.35 & $11.2^a$ & 4.97 & 4.84 & 0.06 & 0.12 & 0.19 & 0.16 \\
\ion{Si}{iv} & Na & $^{bl}$818.15 & $1.3^a$ & 5.07 & 4.93 & 0.15 & 0.24 & 0.33 & 0.25 \\
\ion{C}{iv} & Li & 1548.24 & $212.0^b$ & 5.10 & 4.98 & 0.43 & 1.30 & 1.28 & 0.31 - 0.74 \\
\ion{C}{iv} & Li & 1550.82 & $134.0^b$ & 5.10 & 4.98 & 0.34 & 1.03 & 1.01 & 0.34 \\
\ion{N}{v} & Li & 1238.82 & $81.5^a$ & 5.47 & 5.42 & 0.22 & 0.27 & 0.28 & 0.18 - 0.40 \\
\ion{N}{v} & Li & 1242.80 & $38.5^a$ & 5.47 & 5.42 & 0.23 & 0.28 & 0.30 & 0.17 - 0.40 \\
\ion{S}{vi} & Na & $^{bl}$706.50 & $0.5^a$ & 5.40 & 5.47 & 0.62 & 0.51 & 0.52 &  \\
\ion{S}{vi} & Na & 933.40 & $19.1^a$ & 5.48 & 5.42 & 0.23 & 0.30 & 0.31 & 0.43 - 0.59 \\
\ion{S}{vi} & Na & $^{bl}$944.55 & $9.4^a$ & 5.86 & 5.80 & 0.53 & 0.56 & 0.58 & 0.88 - 1.15 \\
\ion{S}{vi} & Na & 712.68 & $0.9^a$ & 5.57 & 5.51 & 0.33 & 0.39 & 0.40 &  \\
\ion{O}{vi} & Li & 1031.93 & $354.0^a$ & 5.77 & 5.71 & 0.35 & 0.50 & 0.50 & 0.58 - 0.71 \\
\ion{O}{vi} & Li & 1037.64 & $192.0^a$ & 5.77 & 5.71 & 0.32 & 0.46 & 0.46 & 0.41 - 0.76 \\
			\noalign{\smallskip}\hline
			\noalign{\smallskip}
    \multicolumn{10}{p{0.8\linewidth}}{\textbf{Notes.} Same as Table\;\ref{tab:trcorona}.}
		\end{tabular}
	\normalsize
	\label{tab:lilike}
\end{table*}

The predictions for these lines are still a long way from observations relative to other lines, but the ratios are now similar to those of other Li-like and Na-like ions in this work. If the abundance of Si is higher than the photospheric values adopted here, as suggested in Sect.\;\ref{sec:trcorona}, it would bring the ratios for these lines to 0.5-0.6. These are then similar to the \ion{O}{vi} and \ion{S}{vi} results in this section. The alternative observations of \citeauthor{brekke1993} from the QR A region, shown in Tab.\;\ref{tab:lilike}, are similar to \citeauthor{warren2005} for these lines, but those from the \citeauthor{brekke1993} QR region are more than a factor of two lower than \citeauthor{warren2005}. So, the solar variability for these lines is similar to the Li-like \ion{C}{iv} lines, which form around the same temperature; further analysis is required to assess this issue.

The 1128.35\,\AA\ \ion{Si}{iv} line is much further from observations than the resonance lines and the line at 818.15\,\AA. The energy levels which emit the 1128.35\,\AA\ and 818.15\,\AA\ lines are more than two times higher in energy than the upper levels of the resonance lines. As highlighted in Sect.\;\ref{sec:context}, such lines could be affected by time-dependent ionisation or non-Maxwellian electron energy distributions, (although the line at 1128.35\,\AA\ ($3p\; ^2P^o_{3/2} - 3d\; ^2D_{5/2}$) is not strictly a $\Delta n \geq 1$ transition). \citet{pietarila2004} showed that the 818.15\,\AA\ line could be enhanced by a factor of three due to velocity redistribution in the quiet Sun. Here, the ratio for the very weak 818.15\,\AA\ line is similar to the resonance lines, but it has a higher excitation energy than the 1128.35\,\AA\ line. $R_f^{alt}$ from the \cite{parenti2005} observations for all of these \ion{Si}{iv} lines shows a similar trend. The lack of agreement in the ratios for the two lines from highly-excited levels and the ratio for the 818.15\,\AA\ being close to the resonance lines, however, produces conflicting evidence about whether such effects are influencing the lines. An alternative explanation might be that a blend affects the 1128.35\,\AA\ line, and a new work suggests that neutral carbon is making a contribution to the observable intensity of this line (Storey, Dufresne and Del Zanna, in preparation).

The \ion{O}{iv} 1401.15\,\AA\ and \ion{Si}{iv} 1402.77\,\AA\ lines are frequently used for diagnostics using the Interface Region Imaging Spectrometer (IRIS). With the observations here, the 1401.15\,\AA/1402.77\,\AA\ intensity ratio is 0.25, which is close to other works in the quiet Sun, such as \citet{doschek2001}. Using coronal modelling the ratio is 2.25 and with the full model it is 0.47. If the silicon abundance is higher than photospheric, this will bring the ratio using the full models very close to observations. \citet{martinez2016} studied these two lines using two-dimensional, radiative, magneto-hydrodynamical (MHD) models. They found the intensity ratio of the lines differs from the ratio derived assuming steady state equilibrium, making it difficult to assign a single value for density from the ratio. In their models they used the coronal approximation, but the results here highlight that the present atomic models would be better for such an analysis.

\subsection{Lower transition region}
\label{sec:lowertr}

The lower transition region is broadly defined here to be below 100\,000\,K. Discussion of the results for singly-charge ions will be deferred until the next section because many of them are predicted to form in upper chromosphere using the present models. For the \ion{Si}{iii} lines the observations of \citet{pinfield1999} were chosen because they covered all the main lines emitted by this ion; very good consistency was also found for lines within multiplets and within the ion as a whole. For the majority of the lines, the predicted intensities decrease compared to the coronal approximation. This is because charge transfer reduces the peak fractional abundance of \ion{Si}{iii} and suppression of dielectronic recombination from \ion{Si}{iv} narrows the temperature range over which \ion{Si}{iii} forms, as highlighted in \citet{dufresne2021picrm}. Lines from this ion are almost the only case in which the emission measures using the present models move further from those of other ions. Despite this, it is noted that the predicted to observed ratio using the full models are mostly in the range 0.6-0.7, which is remarkably similar to the \ion{Si}{viii} lines and again suggests the FIP bias noted for Si in Sect.\;\ref{sec:uppertr}.

\citet{delzanna2015si3} highlighted that the intensity of the 1206.51\,\AA\ line in \citeauthor{pinfield1999} is high, but the intensity is not as high as \citeauthor{warren2005}, \citeauthor{brekke1993} or \citeauthor{vernazza1978}. The lines in the multiplet around 1298\,\AA\ have similar ratios to each other, except for the 1303.32\,\AA\ line. For the latter line, \citeauthor{pinfield1999} suggested a blend with two \ion{S}{i} lines, but there are no transitions in \textsc{Chianti} to take that into account here. A similar situation applies for the multiplet near 1110\,\AA, where the low ratio for the 1108.37\,\AA\ line is caused by a second order blend with \ion{O}{iv}, which is not taken into account in the predicted intensities.

The increase in the predicted intensity of the \ion{Si}{iii} 1892.03\,\AA\ inter-combination line in the full model is more than a factor of two. The contribution function in Fig.\;\ref{fig:sicontribs} shows how the inclusion of charge transfer allows the line to form at lower temperature. The steep rise in the DEM there more than compensates for the reduction in the peak of the contribution function. The effective temperature has changed from 32\,000\,K to 20\,000\,K, confirming the decrease that \citet{baliunas1980} predicted. The only observation for the line is from \citet{nicolas1977}. The ratio differs from the other \ion{Si}{iii} lines, but the on-disc intensities of \ion{Si}{ii} and \ion{Si}{iii} from \citet{nicolas1977} are 0.56-0.85 times lower than the other works considered here, with the exception of \citet{parenti2005} in some cases. This could account for the inter-combination line ratio being higher than the ratios of the \citet{pinfield1999} lines.

The anomalously high intensity of the 1312.59\,\AA\ line compared to predictions leads to claims that it reflects the presence of non-Maxwellian electron distributions in the plasma because the upper level involved in the transition is higher than the main \ion{Si}{iii} lines, \citep[see for example][]{dufton1984}. The upper level, however, is only 10-20 per cent higher in energy. Since the line also forms at higher temperature, its upper level is not any higher in energy relative to the line formation temperature than the other \ion{Si}{iii} lines. \citet{pinfield1999} claimed that changes to the ion peak abundance and temperature if non-Maxwellian distributions are present are insufficient to alter the line intensity, and so that should not affect the analysis here. \citet{delzanna2015si3} found good agreement between an HRTS observation and predictions for this line when using estimates of the temperature and DEM distribution, as well as improved atomic data. The HRTS observation is around a factor of two lower than the \citeauthor{pinfield1999} intensity from SUMER.

There is an increase by a factor of two in the predicted intensity of the \ion{C}{iii} inter-combination line at 1908.73\,\AA. This is caused by only a small increase in the populations of \ion{C}{iii} at low temperatures through photo-ionisation of \ion{C}{ii}. The observed intensity is estimated from the ratio of this line at 12\arcsec\ inside the white light limb with the \ion{Si}{iii} 1892.03\,\AA\ line given by \citet{mariska1978}. The ratio was multiplied by the disc-centre intensity of the \ion{Si}{iii} line given in Tab.\;\ref{tab:lowertr}. These two lines are close in wavelength, which avoids issues with instrumental efficiency. Using the line ratio from \citet{mariska1978} with the \ion{O}{iv} 1401.15\,\AA\ line gives an intensity within around 20\% of the above estimate. The only changes the modelling has on the remaining \ion{C}{iii} lines is due to density effects on free electron processes. The changes to the ratios for these lines are the same as discussed in \citet{dufresne2019}.

\begin{table*}
	\caption{Comparison of predicted and observed quiet Sun radiances for lower transition region lines.}
	\centering
	\begin{tabular}{p{1.3cm}>{\raggedright}p{0.7cm}>{\raggedleft}p{1.3cm}>{\raggedleft}p{1.3cm}>{\centering}p{0.7cm}>{\raggedright}p{0.8cm}>{\centering}p{0.7cm}>{\centering}p{0.7cm}>{\centering}p{0.7cm}>{\centering\arraybackslash}p{1.9cm}}
			\hline\hline \noalign{\smallskip}
			Ion & Seq & $\lambda_{\rm obs}$ & $I_{\rm obs}$ & $T_{\rm c}$ & $T_{\rm f}$ & $R_{\rm c}$ & $R_{\rm e}$ & $R_{\rm f}$ & $R_{\rm f}^{\rm alt}$ \\
			\noalign {\smallskip} \hline \noalign {\smallskip}

\ion{Si}{iii} & Mg & 1892.03 & $832.0^g$ & 4.50 & 4.30 & 0.47 & 0.85 & 1.08 &  \\
\ion{Si}{iii} & Mg & 1206.51 & $630.0^e$ & 4.60 & 4.49 & 0.69 & 0.88 & 0.69 & 0.37 - 0.81 \\
\ion{Si}{iii} & Mg & 1294.54 & $4.9^e$ & 4.65 & 4.59 & 0.84 & 0.85 & 0.60 & 0.78 - 1.31 \\
\ion{Si}{iii} & Mg & $^{sb}$1298.89 & $16.9^e$ & 4.65 & 4.59 & 0.83 & 0.85 & 0.60 & 0.84 - 0.92 \\
\ion{Si}{iii} & Mg & 1296.73 & $3.8^e$ & 4.68 & 4.63 & 0.83 & 0.85 & 0.60 & 1.36 \\
\ion{Si}{iii} & Mg & 1303.32 & $7.0^e$ & 4.65 & 4.58 & 0.55 & 0.57 & 0.40 &  \\
\ion{Si}{iii} & Mg & 1108.37 & $4.3^e$ & 4.68 & 4.62 & 0.61 & 0.58 & 0.41 &  \\
\ion{Si}{iii} & Mg & $^{sb}$1109.97 & $6.5^e$ & 4.68 & 4.62 & 1.00 & 0.94 & 0.67 & 0.46 \\
\ion{Si}{iii} & Mg & 1113.23 & $15.5^e$ & 4.68 & 4.62 & 1.07 & 0.99 & 0.70 & 0.82 \\
\ion{Si}{iii} & Mg & 1312.59 & $2.7^e$ & 4.71 & 4.66 & 0.62 & 0.51 & 0.36 &  \\
\ion{C}{iii} & Be & 1908.73 & $\approx 220$ & 4.77 & 4.56 & 0.35 & 0.63 & 0.82 &  \\
\ion{C}{iii}* & Be & 977.04 & $702.0^c$ & 4.82 & 4.70 & 0.65 & 0.89 & 0.89 & 0.51 - 1.18 \\
\ion{C}{iii} & Be & 1174.88 & $37.4^c$ & 4.82 & 4.72 & 0.65 & 0.85 & 0.85 & 0.57 - 0.73 \\
\ion{C}{iii}* & Be & 1175.74 & $104.0^c$ & 4.82 & 4.72 & 0.71 & 0.89 & 0.89 & 0.53 - 0.59 \\
\ion{C}{iii} & Be & 1176.37 & $36.2^c$ & 4.82 & 4.72 & 0.67 & 0.87 & 0.87 & 0.58 - 0.60 \\
\ion{C}{iii} & Be & 1247.40 & $2.1^b$ & 4.88 & 4.79 & 0.76 & 0.71 & 0.72 & 0.74 \\
\ion{O}{iii} & C & 1660.80 & $19.3^b$ & 4.87 & 4.72 & 0.44 & 0.63 & 0.81 & 1.09 \\
\ion{O}{iii}* & C & $^{sb}$833.74 & $51.0^a$ & 4.95 & 4.85 & 0.98 & 1.09 & 1.13 & 1.10 - 2.45 \\
\ion{O}{iii} & C & 835.10 & $11.7^a$ & 4.95 & 4.85 & 1.08 & 1.21 & 1.25 &  1.13 - 2.07 \\
\ion{O}{iii}* & C & 835.26 & $78.6^a$ & 4.95 & 4.85 & 0.89 & 0.99 & 1.03 & 1.27 - 2.61 \\
\ion{Ne}{iii}* & O & $^{sb}$489.40 & $4.3^f$ & 5.14 & 5.05 & 1.58 & 1.10 & 1.18 &  \\
			\noalign{\smallskip}\hline
			\noalign{\smallskip}
    \multicolumn{10}{p{0.8\linewidth}}{\textbf{Notes.} Same as Table\;\ref{tab:trcorona}.}
    \end{tabular}
	\normalsize
	\label{tab:lowertr}
\end{table*}

Photo-ionisation and density effects have also had a considerable impact on the \ion{O}{iii} 1660.80\,\AA\ inter-combination line. In \citet{dufresne2021pico}, it is highlighted how photo-ionisation and charge transfer oppose each other in the formation of \ion{O}{ii} and \ion{O}{iii}, with charge transfer reducing the fractional population of \ion{O}{iii} that photo-ionisation creates at low temperatures. The overall change to the ion balance when the two processes are combined in the collisional-radiative model depends on pressure, neutral hydrogen abundance and the strength of photo-ionising radiation. The better agreement in ratios for the 1660.80\,\AA\ line and the higher temperature \ion{O}{ii} lines (shown in the next section) suggests that the balance of these processes in the models is about right. The effective temperature of the inter-combination line has lowered from 74100\,K to 52500\,K using the new modelling, which is one of the largest changes of all the oxygen lines. The enhancement in the \ion{O}{iii} populations at lower temperature only causes small changes in the other \ion{O}{iii} lines, and they all show very good agreement, as shown both Tab.\;\ref{tab:lowertr} and Tab.\;\ref{tab:applowertr}.

Observations from the low charge states of neon are hard to find, making it difficult to check the significant changes seen in ion formation from PI shown for these ions in \citet{dufresne2021picrm}. The intensities of the \ion{Ne}{iii} multiplet around 489.4\,\AA, which are blended in the Skylab observations, decrease notably in the electron collisional model. They are slightly enhanced compared to this in the full model because of the significant presence of \ion{Ne}{iii} at lower temperatures due to the PI of \ion{Ne}{ii}.

Trends in isoelectronic sequence can be checked again by comparing Be-like ions. Taking, for instance, the inter-combination lines, the \ion{C}{iii} line is altered by a factor of two, the \ion{N}{iv} line by 25 per cent and the \ion{O}{v} line is not affected at all. The allowed lines in \ion{C}{iii} are enhanced by 30-40 per cent, while the same lines in \ion{N}{iv} and \ion{O}{v} are unchanged. The lines are all now in good agreement with observations using the full models, except for the \ion{O}{v} inter-combination line.

\subsection{Transition region-chromosphere boundary}
\label{sec:chromotr}

\begin{table*}
	\caption{Comparison of predicted and observed quiet Sun radiances for transition region-chromosphere boundary lines.}
	\centering
	\begin{tabular}{p{1.3cm}>{\raggedright}p{0.7cm}>{\raggedleft}p{1.3cm}>{\raggedleft}p{1.3cm}>{\centering}p{0.7cm}>{\raggedright}p{0.8cm}>{\centering}p{0.7cm}>{\centering}p{0.7cm}>{\centering}p{0.7cm}>{\centering\arraybackslash}p{1.9cm}}
			\hline\hline \noalign{\smallskip}
			Ion & Seq & $\lambda_{\rm obs}$ & $I_{\rm obs}$ & $T_{\rm c}$ & $T_{\rm f}$ & $R_{\rm c}$ & $R_{\rm e}$ & $R_{\rm f}$ & $R_{\rm f}^{\rm alt}$ \\
			\noalign {\smallskip} \hline \noalign {\smallskip}

		\ion{C}{ii} & B & 2325.40 & $\approx 210$ & 4.33 & 4.03 & 0.36 & 0.55 & 2.08 & \\
		\ion{C}{ii}* & B & 1334.53 & $937.0^a$ & 4.40 & 4.17 & 0.56 & 0.66 & 0.94 & 1.31 - 1.66 \\
		\ion{C}{ii}* & B & 1335.71 & $1350.0^a$ & 4.40 & 4.17 & 0.75 & 0.90 & 1.29 & 1.45 - 2.25 \\
		\ion{C}{ii}* & B & 1036.34 & $57.9^a$ & 4.48 & 4.34 & 0.92 & 0.80 & 0.68 & 1.11 - 1.21 \\
		\ion{C}{ii}* & B & 1037.00 & $70.1^a$ & 4.48 & 4.34 & 1.50 & 1.31 & 1.12 & 1.75 - 1.96 \\
		\ion{C}{ii} & B & 903.99 & $10.0^a$ & 4.51 & 4.40 & 2.22 & 1.67 & 1.28 & 1.76 - 4.07 \\
		\ion{C}{ii} & B & 904.46 & $6.3^a$ & 4.51 & 4.40 & 1.78 & 1.34 & 1.02 & 1.38 - 3.03 \\
		\ion{C}{ii} & B & 903.59 & $9.2^a$ & 4.52 & 4.40 & 1.62 & 1.21 & 0.93 & 1.46 - 2.08 \\
		\ion{C}{ii} & B & 904.14 & $23.1^a$ & 4.52 & 4.39 & 3.21 & 2.40 & 1.85 & 3.15 - 4.26 \\
		\ion{C}{ii} & B & $^{sb}$1323.91 & $1.7^b$ & 4.58 & 4.48 & 2.47 & 1.30 & 0.97 & 0.74 \\
		\ion{O}{ii} & N & 832.75 & $11.2^a$ & 4.57 & 4.51 & 3.02 & 2.67 & 2.20 & 1.91 - 4.42 \\
		\ion{O}{ii} & N & 834.45 & $35.5^a$ & 4.57 & 4.51 & 2.85 & 2.52 & 2.08 & 2.38 - 3.81 \\
		\ion{O}{ii} & N & 833.32 & $24.5^a$ & 4.58 & 4.51 & 2.75 & 2.43 & 2.01 & 2.25 - 3.59 \\
		\ion{O}{ii} & N & $^{sb}$718.49 & $15.0^a$ & 4.67 & 4.62 & 2.10 & 1.39 & 1.22 & 1.25 \\
		\ion{O}{ii} & N & $^{bl}$796.66 & $2.8^d$ & 4.71 & 4.67 & 1.71 & 1.34 & 1.23 &  \\
		\noalign{\smallskip}\hline
		\noalign{\smallskip}
		\multicolumn{10}{p{0.8\linewidth}}{\textbf{Notes.} Same as Table\;\ref{tab:trcorona}.}
	\end{tabular}
	\normalsize
	\label{tab:trchromo}
\end{table*}

While line emission for neutrals as well as singly-charged ions from low FIP elements has long been known to be affected by opacity and dynamical effects in the chromosphere, \citep[see for example][]{skelton1982, lanzafame1994}, it has been shown more recently to apply to \ion{C}{ii}, an higher FIP element, by \citet{rathore2015}. They highlighted that, even in ionisation equilibrium, photo-ionisation of \ion{C}{i} causes \ion{C}{ii} to form in the upper chromosphere, instead of the lower transition region predicted by the coronal approximation of \textsc{Chianti}. Their radiation hydrodynamical modelling predicts the \ion{C}{ii} 1335.7\,\AA /1334.5\,\AA\ intensity ratio to be in the range 1.4-1.7, which is closer to observations than the optically thin value of two in the high density limit.

In \citet{dufresne2021picrm}, the models for higher FIP elements C, N, O, and Ne confirm that the singly-charged ions are all shifted to the upper chromosphere when the new atomic processes are added to the coronal approximation. Radiative transfer and dynamical effects should then become more important in both the formation of the ions and their emission lines; this has not been taken into account in the present models. The results are also limited by the DEM and contribution functions being cut off at ${\rm log}\,T=3.9$, whereas many of the lines could form below that temperature according to the current models. These are all limitations in benchmarking the new models against observations for ions in this region.

The \ion{C}{ii} inter-combination line at 2325.40\,\AA\ is enhanced by almost a factor of six. This increase results from higher density in this region and the presence of \ion{C}{ii} at lower temperatures due to photo-ionisation of \ion{C}{i}. The observation is estimated from the \citet{mariska1978} ratio with the \ion{O}{iv} 1401.15\,\AA\ line, but this estimate is the most uncertain in this work because the ratio is taken from 2\arcsec\ above the limb and there are no other observations to test it against. Since the two observed, \ion{C}{ii} resonance lines at 1334.53\,\AA\ and 1335.71\,\AA\ (the latter is a self-blend) form at lower temperature than the other dipole-allowed lines from this ion, their predicted intensities are enhanced for the same reason as the inter-combination line from this ion. The doublet with wavelengths at 1036.34\,\AA\ and 1037.00\,\AA\ forms at temperatures above the peak in ion formation, by contrast. This means their intensities are reduced not only by the shift of ion formation to lower temperature when density effects are included, but also because photo-ionisation of \ion{C}{ii} reduces the peak ion abundance.

It is not surprising that the ratios within each of the above multiplets are not similar to each other, compared to the results seen in previous sections. \citet{doschek1999} mentioned that the 1037.00\,\AA\ line has both a stronger decay rate and an higher population of the lower level involved in the transition, relative to the weaker 1036.34\,\AA\ line. This effect, known as cross frequency redistribution \citep[see][for example]{koncewicz2007}, causes intensity to be transferred from the stronger line to the weaker line. It is, therefore, understandable that the observed intensities of the stronger lines in each of these multiplets are less than predictions from optically thin modelling, while the weaker lines are greater than predictions.

Considering the \ion{C}{ii} lines around 904\,\AA, \citet{parenti2019} noted that theoretical intensities for the lines were significantly higher than observations in the QS. These are weak lines and should be less prone to opacity effects than the stronger lines emitted by this ion. However, because the discrepancies are even greater in prominences, \citeauthor{parenti2019} highlighted that hydrogen is likely to be absorbing photons from the lines, given how close they form to the edge of the Lyman continuum. Here, the ratios in the full model are all much closer to the other lines emitted by this ion, except for the strongest line at 904.14\,\AA. This indicates that redistribution of intensity from this line could be taking place. \citet{parenti2019} stated that the weak line at 1323.91\,\AA, which is less prone to opacity effects, may be a better line from \ion{C}{ii} to constrain the DEM at lower temperatures. This is a self-blend of four lines involving transitions from highly excited levels ($2s\,2p^2\;^2D - 2p^3\;^2D^o$), and they could be prone to time dependent ionisation and non-Maxwellian electrons. However, it is seen here how the full model brings emission for this multiplet into agreement with the other \ion{C}{ii} lines emitted from less excited levels.

\citet{dufresne2020} showed how the \ion{O}{ii} lines lie closer to the emission measure of other lines when using the electron collisional models, but the predictions were still a factor of two from observations. Further improvements can now be seen when using the full models, as shown in Tab.\;\ref{tab:trchromo}. When using the full model, the \ion{O}{ii} line at 718.49\,\AA, which is observed by the Spectral Imaging of the Coronal Environment (SPICE) spectrometer \citep{anderson2020} on board Solar Orbiter, now shows excellent agreement with lines used to fit the DEM at this temperature range. The improvement primarily comes from photo-ionisation reducing the peak in the fractional population of \ion{O}{ii}, as shown in the collisional-radiative model of \citet{dufresne2021pico}. It appears that the atomic processes included here will be required to interpret the data from SPICE for this line. The 796.66\,\AA\ line is identified by \citet{parenti2005} as being emitted from \ion{O}{ii}. The contribution from this ion is a self-blend, plus there is a slightly stronger contribution from \ion{S}{iii} (52 per cent of the total intensity). The blend with \ion{S}{iii} explains why there is a slightly smaller change in predictions compared to the 718.49\,\AA\ line, which forms at a similar temperature.

Better agreement is also achieved using the full models for the \ion{O}{ii} lines emitted around 833\,\AA. The emission measures of these lines, though, are still far from those of neighbouring lines, although there is consistency in the ratios within the multiplet. Hydrogen absorption might, as with the \ion{C}{ii} 904.14\,\AA\ line, account for the reduction in observed intensities. However, these lines do not form as low in the atmosphere as the 904\,\AA\ lines and are further away from the hydrogen Lyman edge. The \ion{O}{ii} 833\,\AA\ lines are also very close in wavelength to the lines emitted by \ion{O}{iii}. In all of the models tested here, the predictions for the \ion{O}{iii} 833\,\AA\ lines reflect observations well, which highlights there is not a problem of resolving blends in the observations, \citep[see][Fig. 4]{warren2005}.

The results show that emission from \ion{N}{ii} is little affected by the atomic processes added to the coronal approximation; the results are discussed in the Appendix, Sect.\;\ref{sec:apptrchromo}. The collisional-radiative model for \ion{S}{ii} shows that it begins forming very low down in the chromosphere through photo-ionisation and charge transfer ionisation of \ion{S}{i}. Like \ion{Si}{ii}, which shows the least consistency in predictions compared to observations in this work, it will require hydrodynamic and/or radiative transfer calculations to fully assess its line formation. Results for both these ions are also given in Sect.\;\ref{sec:apptrchromo} of the Appendix.

Looking at trends in isoelectronic sequence again, there are many lines in the B-like sequence for comparison. All the allowed lines from this sequence are in good agreement with observations as a whole. The greatest changes are at low temperatures, particularly with the \ion{C}{ii} inter-combination line and lines at 904\,\AA; there are no equivalent changes for \ion{N}{iii} and \ion{O}{iv}. In this sequence, predicted intensities for \ion{Mg}{viii} also change little; the only discrepancy coming from a likely FIP bias. The same applies to the C-like sequence, where all lines are relatively unaffected, except for \ion{N}{ii}. Taking Al-like ions as an example of a third-row sequence, it is clear that \ion{S}{iv} lines are not affected by the same issues as those from \ion{Si}{ii}. Similarly, it appears to be a FIP bias and inter-combination line issues that account for the discrepancy in the Mg-like sequence between \ion{Si}{iii} and \ion{S}{v}. Thus, apart from the special case of the Li- and Na-like sequences, it appears that the region of formation (and emission measure) is a greater influence on line emission than isoelectronic sequence.

\section{Conclusions}
\label{sec:concl}

The most significant changes to total line intensities brought about by the present atomic models for the transition region occur at temperatures below 100\,000\,K. The DEM is steep in that region and so the intensities alter significantly compared to the coronal approximation. Overall, the present models bring improved consistency within ions and between neighbouring ions. Density effects shifting ion formation to lower temperatures causes the intensities of lower temperature lines to increase relative to higher temperature lines within the same ion. In all cases, predictions from the new models have improved the agreement with observations, with the exception of the \ion{S}{ii} lines around 1255\,\AA\ (see Sect.\;\ref{sec:apptrchromo}) and the \ion{Si}{iii} lines. For the latter ion, however, the ratios are similar to the \ion{Si}{viii} lines, and may indicate a FIP bias for this element. 

In many cases, the greatest changes are produced by the new atomic processes added to the coronal approximation: photo-ionisation and charge transfer. The influence of these processes brings much better agreement for the \ion{O}{ii} 718.49\,\AA\ and 796.66\,\AA\ lines, and the \ion{O}{iii} 1660.80\,\AA\ and \ion{Si}{iii} 1892.03\,\AA\ inter-combination lines. They all show factors of two changes in the predicted intensities, but even greater changes occur in this temperature range for other inter-combination lines and Li- and Na-like lines. One of the most notable is \ion{Si}{iv}, for which predicted intensities increase by a factor of six.

In the region 100\,000-300\,000\,K the DEM is relatively flat. Therefore, the shifts in ion formation caused by density on the electron collisional processes do not alter the resulting line intensities as much. Despite this, there are still significant changes in ion formation in this region. Consequently, using these types of models for line ratio diagnostics in isothermal conditions, such as in active regions and flares, could produce notable changes. The intensities which show the greatest change in this region are inter-combination and Li- and Na-like lines, which are all enhanced by 20-40 per cent using the present models. These changes bring improved consistency in the emission measures of the lines, but they cannot account fully for the emission from \ion{N}{v} and \ion{S}{vi}. Above 300\,000\,K, the changes to intensities are in the range of 10-20 per cent, except for Li-like \ion{O}{vi}.

Over the majority of the transition region, the present atomic models are an improvement on the coronal approximation. They seem to capture the relevant atomic processes for emission in this region and the upper chromosphere. This straightforward comparison with observations clearly indicates which ions are more affected by the new atomic models, and that the included atomic processes are contributing to the observed lines. The models appear to be important for interpreting emission observed by such important missions as SPICE, which observes all the transition region ions of oxygen, and IRIS, which observes lines from \ion{C}{ii}, \ion{O}{iv}, \ion{Si}{iv} and \ion{S}{iv}. These lines have all shown significant changes here.

As highlighted at various points in this work, the present atomic models are the beginning point from which to add the many other important effects needed to model line emission from the TR. Radiative transfer effects are clearly important in the chromosphere and lower TR just to interpret disc-centre, averaged intensities \citep{pietarila2004}, let alone for centre-to-limb variation and irradiance measurements of the whole Sun and other stars. \citet{hansteen1993} and \citet{olluri2013} have also shown, for example, the importance of time dependent ionisation for upper TR emission, even in the quiet Sun, where ubiquitous redshifts and departures from ionisation equilibrium intensity ratios are routinely observed. In addition, such effects are required to explain the variability present not only in temporally- and spatially-resolved observations, but also in the average  intensities used here. For Li- and Na-like ions, all of these effects, and perhaps more, might be required to correctly interpret their emission. The present work has validated not only the use of these types of atomic models in such modelling, but it has also highlighted which lines may require more advanced modelling to correctly interpret their emission.

\section{Data availability}

Ion fractions from the atomic models for the transition region have been made available at the CDS via anonymous ftp to cdsarc.u-strasbg.fr (130.79.128.5) or via http://cdsarc.u-strasbg.fr/viz-bin/qcat?J/MNRAS. All the electron impact excitation and radiative decay data required to calculate level populations were obtained from the \textsc{Chianti} v.10 atomic database \citet{delzanna2021v10}, as well as the coronal approximation ion fractions. Solar observations were obtained from published data in the referenced journal articles.

\section*{acknowledgements}

Acknowledgement is given to the referee, whose comments have greatly helped to improve the quality and focus of the text.

RPD acknowledges studentships from the STFC (UK) Doctoral Training Programme, the University of Cambridge Isaac Newton Trust and the Cambridge Philosophical Society. GDZ acknowledges support from STFC (UK) via the consolidated grants to the atomic astrophysics group at the Department of Applied Mathematics and Theoretical Physics, University of Cambridge (ST/P000665/1 and ST/T000481/1). \
	
Most of the atomic rates used in the present study were produced by the UK APAP network, funded by STFC via several grants to the University of Strathclyde. \
	
\textsc{Chianti} is a collaborative project involving George Mason University, the University of Michigan, the NASA Goddard Space Flight Centre (USA) and the University of Cambridge (UK). \

\bibliographystyle{mnras}

\bibliography{pi_obs}

\appendix

\section{Results for other lines}

\begin{table}
	\caption{List of lines used to constrain the DEM at the lowest and highest temperatures.}
	\centering
	\begin{tabular}{p{0.5cm}>{\raggedleft}p{1.3cm}>{\raggedleft}p{0.9cm}>{\centering}p{0.4cm}>{\raggedright}p{0.4cm}>{\centering}p{0.4cm}>{\centering}p{0.4cm}>{\centering\arraybackslash}p{0.4cm}}
			\hline\hline \noalign{\smallskip}
			Ion & $\lambda_{\rm obs}$ & $I_{\rm obs}$ & $T_{\rm c}$ & $T_{\rm f}$ & $R_{\rm c}$ & $R_{\rm e}$ & $R_{\rm f}$ \\
			\noalign {\smallskip} \hline \noalign {\smallskip}

\ion{C}{i} & $^{sb}1656.98$ & $284.0^b$ & 4.02 & 3.96 & 0.75 & 0.78 & 0.69 \\
\ion{C}{i} & $^{bl}$1560.31 & $96.2^b$ & 4.04 & 3.98 & 0.68 & 0.72 & 0.62 \\
\ion{C}{i} & $^{sb}$1560.69 & $114.0^b$ & 4.04 & 3.97 & 1.76 & 1.83 & 1.49 \\
\ion{N}{i} & 1199.57 & $20.6^d$ & 4.16 & 4.03 & 0.44 & 0.41 & 0.33 \\
\ion{N}{i} & 1200.23 & $13.4^d$ & 4.16 & 4.03 & 0.46 & 0.42 & 0.34 \\
\ion{Fe}{xi} & 352.67 & $30.5^a$ & 6.11 & 6.11 & 1.18 & 1.18 & 1.17 \\
\ion{Si}{x} & 347.40 & $44.7^a$ & 6.12 & 6.12 & 1.02 & 1.03 & 1.03 \\
\ion{Si}{x} & $^{sb}$356.03 & $24.2^a$ & 6.12 & 6.11 & 0.97 & 0.99 & 0.98 \\
\ion{Fe}{xii} & 364.45 & $34.1^a$ & 6.14 & 6.14 & 0.95 & 0.93 & 0.93 \\
\ion{Si}{xi} & 303.34 & $125.0^a$ & 6.15 & 6.15 & 0.80 & 0.82 & 0.82 \\
\ion{Si}{xii} & 520.68 & $5.0^a$ & 6.18 & 6.17 & 1.13 & 1.12 & 1.13 \\
\noalign{\smallskip}\hline
			\noalign{\smallskip}
    \multicolumn{8}{p{0.95\linewidth}}{\textbf{Notes.} Ion - principal emitting ion; observed wavelength $\lambda_{\rm obs}$ (\AA), where superscript `sb' denotes a self-blend and `bl' a blend; the measured radiance $I_{\rm obs}$ (ergs cm$^{-2}$ s$^{-1}$ sr$^{-1}$) using: a) \citeauthor{warren2005}, b) \citeauthor{brekke1993}, c) \citeauthor{wilhelm1998}, d) \citeauthor{parenti2005}, e) \citeauthor{pinfield1999}, f) \citeauthor{vernazza1978}, g) \citeauthor{nicolas1977}, and h) \citeauthor{andretta2014}; $T$ - the effective temperature for each line (logarithmic values, in K), and $R$ - the ratio between the predicted and observed intensities; subscripts of $T$ and $R$ refer to results obtained using: c) \textsc{Chianti} coronal approximation ion fractions, e) ion fractions from electron collisional models, and f) ion fractions from full models.}
		\end{tabular}
	\normalsize
	\label{tab:demlines}
\end{table}

\begin{table*}
	\caption{Comparison of predicted and observed quiet Sun radiances for other upper transition region lines.}
	\centering
	\begin{tabular}{p{1.3cm}>{\raggedright}p{0.7cm}>{\raggedleft}p{1.3cm}>{\raggedleft}p{1.3cm}>{\centering}p{0.7cm}>{\raggedright}p{0.8cm}>{\centering}p{0.7cm}>{\centering}p{0.7cm}>{\centering}p{0.7cm}>{\centering\arraybackslash}p{1.9cm}}
			\hline\hline \noalign{\smallskip}
			Ion & Seq & $\lambda_{\rm obs}$ & $I_{\rm obs}$ & $T_{\rm c}$ & $T_{\rm f}$ & $R_{\rm c}$ & $R_{\rm e}$ & $R_{\rm f}$ & $R_{\rm f}^{\rm alt}$ \\
			\noalign {\smallskip} \hline \noalign {\smallskip}

\ion{S}{iv}* & Al & 1072.98 & $7.6^d$ & 5.01 & 4.90 & 0.66 & 0.71 & 0.74 &  0.56 \\
\ion{S}{iv}* & Al & 1062.75 & $3.7^d$ & 5.02 & 4.90 & 0.87 & 0.92 & 0.95 &  \\
\ion{S}{iv}* & Al & 750.22 & $6.9^a$ & 5.05 & 4.93 & 0.84 & 0.76 & 0.77 &  \\
\ion{S}{iv} & Al & 753.74 & $1.5^a$ & 5.05 & 4.93 & 0.83 & 0.76 & 0.77 &  0.80 \\
\ion{S}{iv} & Al & 748.40 & $2.9^a$ & 5.06 & 4.95 & 0.80 & 0.72 & 0.73 & \\
\ion{S}{iv} & Al & $^{sb}$661.40 & $6.9^f$ & 5.12 & 5.03 & 1.01 & 0.86 & 0.86 & \\
\ion{S}{v}* & Mg & 786.47 & $32.1^a$ & 5.21 & 5.13 & 0.95 & 1.03 & 1.00 & 1.11 \\
\ion{N}{iv}* & Be & 765.15 & $80.7^a$ & 5.16 & 5.09 & 0.92 & 0.95 & 0.92 &  \\
\ion{O}{iv}* & B & 787.72 & $58.5^a$ & 5.21 & 5.13 & 1.19 & 1.16 & 1.16 &  1.29 \\
\ion{O}{iv}* & B & 790.19 & $108.0^a$ & 5.21 & 5.13 & 1.30 & 1.26 & 1.25 &  \\
\ion{O}{iv} & B & 554.10 & $40.1^a$ & 5.22 & 5.15 & 1.23 & 1.05 & 1.05 &  0.26 \\
\ion{O}{iv}* & B & 554.55 & $112.0^a$ & 5.23 & 5.16 & 1.10 & 0.94 & 0.94 &  \\
\ion{O}{iv}* & B & 608.38 & $17.7^a$ & 5.24 & 5.18 & 1.18 & 1.05 & 1.05 &  \\
\ion{O}{v} & Be & 758.68 & $6.1^a$ & 5.37 & 5.35 & 1.01 & 0.92 & 0.96 & \\
\ion{O}{v}* & Be & 760.43 & $18.8^a$ & 5.37 & 5.35 & 0.97 & 0.88 & 0.92 &  \\
\ion{O}{v} & Be & 761.99 & $6.3^a$ & 5.37 & 5.35 & 0.94 & 0.86 & 0.89 & \\
\ion{O}{v}* & Be & 629.78 & $338.0^a$ & 5.38 & 5.35 & 1.05 & 0.95 & 0.99 & 1.00 - 1.23 \\
\ion{O}{v} & Be & 1371.34 & $4.1^b$ & 5.38 & 5.36 & 1.07 & 0.96 & 0.99 & \\
\ion{Ne}{iv} & N & 543.91 & $8.3^a$ & 5.28 & 5.21 & 0.81 & 0.76 & 0.78 & 0.71 \\
\ion{Ne}{iv}* & N & 541.14 & $2.4^a$ & 5.29 & 5.23 & 0.93 & 0.88 & 0.91 &  \\
\ion{Ne}{iv}* & N & 542.10 & $4.6^a$ & 5.30 & 5.23 & 0.96 & 0.91 & 0.94 &  \\
\ion{Ne}{v}* & C & $^{sb}$569.82 & $5.4^a$ & 5.46 & 5.44 & 1.02 & 1.10 & 1.10 & \\
\ion{Ne}{v}* & C & 572.31 & $8.8^a$ & 5.47 & 5.44 & 0.86 & 0.93 & 0.94 & 0.94 \\
\ion{Ne}{vi}* & B & 562.81 & $15.6^a$ & 5.65 & 5.61 & 1.03 & 1.07 & 1.02 & 0.90 \\
\ion{Ne}{vi}* & B & $^{bl}$558.61 & $9.6^a$ & 5.67 & 5.63 & 0.99 & 1.01 & 0.97 & \\
\ion{Ne}{vii}* & Be & 895.17 & $4.5^a$ & 5.79 & 5.76 & 0.90 & 0.93 & 0.93 & \\
\ion{Ne}{vii}* & Be & 561.73 & $3.1^a$ & 5.80 & 5.78 & 1.08 & 1.07 & 1.07 & \\
\ion{Ne}{vii} & Be & 465.20 & $120.0^f$ & 5.81 & 5.78 & 0.64 & 0.63 & 0.63 & \\
\ion{Ne}{vii} & Be & 564.61 & $2.2^a$ & 5.81 & 5.78 & 0.49 & 0.48 & 0.49 &  \\
\ion{Ne}{viii}* & Li & 780.30 & $36.8^a$ & 5.99 & 5.99 & 0.97 & 0.93 & 0.96 & \\
\ion{Ne}{viii}* & Li & 770.42 & $73.1^a$ & 6.00 & 5.99 & 0.98 & 0.95 & 0.97 & \\
			\noalign{\smallskip}\hline
			\noalign{\smallskip}
    \multicolumn{10}{p{0.8\linewidth}}{\textbf{Notes.} Ion - principal emitting ion, and `*' denotes a line used to fit the DEM; observed wavelength $\lambda_{\rm obs}$ (\AA), where superscript `sb' denotes a self-blend and `bl' a blend; the measured radiance $I_{\rm obs}$ (ergs cm$^{-2}$ s$^{-1}$ sr$^{-1}$) using: a) \citeauthor{warren2005}, b) \citeauthor{brekke1993}, c) \citeauthor{wilhelm1998}, d) \citeauthor{parenti2005}, e) \citeauthor{pinfield1999}, f) \citeauthor{vernazza1978}, g) \citeauthor{nicolas1977}, and h) \citeauthor{andretta2014}; $T$ - the effective temperature for each line (logarithmic values, in K); $R$ - the ratio between the predicted and observed intensities; subscripts of $T$ and $R$ refer to results obtained using: c) \textsc{Chianti} coronal approximation ion fractions, e) ion fractions from electron collisional models, and f) ion fractions from full models; and, $R^{\rm alt}_{\rm f}$ - the range of $R_{\rm f}$ using the highest and lowest observations from the other sources.}
		\end{tabular}
	\normalsize
	\label{tab:appuppertr}
\end{table*}

\subsection{Other upper transition region lines}
\label{sec:appuppertr}

As noted in Sect.\;\ref{sec:uppertr}, the DEM is relatively flat above 100\,000\,K. So, the shift in ion formation to lower temperature and changes in peak ion abundance seen in the new atomic models does not affect the integrated intensities for many of the lines forming in this part of the atmosphere. This can be seen for the ratios for the S, N and O lines in Tab.\;\ref{tab:appuppertr}, as well as for the \ion{Ne}{iv} lines. Up to and including the formation temperature of \ion{Ne}{iv}, lines from other high FIP elements were used to fit the DEM. The overall agreement in the Ne ratios with other high FIP elements up to these temperatures suggest that its atmospheric abundance is similar to the photospheric value.

For higher charge states of Ne no other lines are used to fit the DEM, and the DEM will adjust so that the predicted to observed ratios will be close to unity for the higher charge Ne lines, regardless of the atomic model being used. The only discrepancies in the ratios are the \ion{Ne}{vii} lines at 465.20\,\AA\ and 564.61\,\AA. The latter line is emitted from the same term as the 561.73\,\AA\ line, which is well-represented by the models, suggesting it is perhaps affected by a blend.

\subsection{Other lower transition region lines}
\label{sec:applowertr}

Photo-ionisation of \ion{N}{ii} produces a smaller enhancement of \ion{N}{iii} at low temperatures compared to the changes seen for \ion{O}{iii}, and the change even in the \ion{N}{iii} inter-combination line is small. \ion{N}{iii} is also depleted to a small degree by charge transfer relative to the electron collisional model, which explains the small decrease in predicted intensity using the full model for this line. The observed intensity was estimated from the \citet{doschek1976} ratio with the \ion{O}{iv} 1401.15\,\AA\ line just inside the limb. However, this is the only observation found for this line and it appears there may have been difficulties in separating the line from blends and the continuum. There are small changes in the remaining lines emitted by \ion{N}{iii} and the results for those are all in good agreement with observations, as shown in Tab.\;\ref{tab:applowertr}. For the \ion{N}{iii} doublet around 991\,\AA, the observations from \citet{parenti2005} are almost a factor of two weaker than \citet{warren2005}, but \citet{vernazza1978} confirms the intensity for the 991.59\,\AA\ line given by \citeauthor{warren2005}.

The contribution function of the \ion{S}{iii} line at 1200.96\,\AA, illustrated in \citet{dufresne2021picrm}, shows a small enhancement at lower temperature due to photo-ionisation of \ion{S}{ii}. This translates into an increase in intensity of 24 per cent for this resonance line. \citet{parenti2005} indicates the other line from the multiplet, at 1194.07\,\AA, has a second order blend from \ion{Ca}{viii} in SUMER. The blend is not included in the predicted intensity, explaining why the prediction for this line is slightly further from observation. Overall, there is better consistency in the ratios for the \ion{S}{iii} lines with the full model, and they all come closer to unity.

\subsection{Other lines from the transition region-chromosphere boundary}
\label{sec:apptrchromo}

The \ion{N}{ii} multiplet around 1085\,\AA\ shows only small increases in predicted intensities with the new models; the results are given in Tab.\;\ref{tab:apptrchromo}. This change occurs more for the electron collisional model, and so it is the shift in ion formation due to density effects that makes the difference. The higher population of \ion{N}{ii} at lower temperature through CT ionisation and PI of \ion{N}{i} has not affected these lines. The effective temperatures of the lines drop from 31\,600\,K to 24\,000\,K.

\begin{table*}
	\caption{Comparison of predicted and observed quiet Sun radiances for other lower transition region lines.}
	\centering
	\begin{tabular}{p{1.3cm}>{\raggedright}p{0.7cm}>{\raggedleft}p{1.3cm}>{\raggedleft}p{1.3cm}>{\centering}p{0.7cm}>{\raggedright}p{0.8cm}>{\centering}p{0.7cm}>{\centering}p{0.7cm}>{\centering}p{0.7cm}>{\centering\arraybackslash}p{1.9cm}}
			\hline\hline \noalign{\smallskip}
			Ion & Seq & $\lambda_{\rm obs}$ & $I_{\rm obs}$ & $T_{\rm c}$ & $T_{\rm f}$ & $R_{\rm c}$ & $R_{\rm e}$ & $R_{\rm f}$ & $R_{\rm f}^{\rm alt}$ \\
			\noalign {\smallskip} \hline \noalign {\smallskip}

\ion{N}{iii} & B & 1749.67 & $\approx 14$ & 4.82 & 4.77 & 0.61 & 0.76 & 0.69 & \\
\ion{N}{iii}* & B & 989.82 & $23.0^a$ & 4.83 & 4.76 & 0.84 & 0.88 & 0.88 & 1.73 \\
\ion{N}{iii}* & B & 991.59 & $44.2^a$ & 4.88 & 4.80 & 0.76 & 0.82 & 0.83 & 0.76 - 1.57 \\
\ion{N}{iii} & B & 764.36 & $5.9^a$ & 4.90 & 4.82 & 1.38 & 1.32 & 1.30 &  \\
\ion{N}{iii}* & B & 685.50 & $6.4^a$ & 4.93 & 4.85 & 1.17 & 1.08 & 1.09 &  \\
\ion{N}{iii}* & B & 685.79 & $15.8^a$ & 4.94 & 4.86 & 1.20 & 1.12 & 1.12 & 0.64 - 0.73 \\
\ion{S}{iii}* & Si & 1194.07 & $7.0^d$ & 4.67 & 4.59 & 0.64 & 0.83 & 0.79 & \\
\ion{S}{iii}* & Si & 1200.96 & $8.0^d$ & 4.67 & 4.59 & 0.88 & 1.15 & 1.09 &  \\
\ion{S}{iii} & Si & 1077.14 & $3.7^a$ & 4.72 & 4.65 & 1.12 & 1.28 & 1.20 & 1.77 \\
\ion{S}{iii} & Si & 680.70 & $2.0^a$ & 4.87 & 4.81 & 1.38 & 1.24 & 1.20 & \\
\ion{O}{iii} & C & $^{sb}$702.89 & $27.5^a$ & 4.97 & 4.87 & 1.16 & 1.18 & 1.20 & \\
\ion{O}{iii}* & C & $^{sb}$703.87 & $43.5^a$ & 4.97 & 4.88 & 1.22 & 1.24 & 1.26 & \\
\ion{O}{iii} & C & 599.56 & $35.7^a$ & 5.01 & 4.91 & 1.03 & 0.92 & 0.92 &   \\
\ion{O}{iii} & C & 525.83 & $17.8^a$ & 5.02 & 4.93 & 1.00 & 0.84 & 0.83 &  0.62 \\
			\noalign{\smallskip}\hline
			\noalign{\smallskip}
    \multicolumn{10}{p{0.8\linewidth}}{\textbf{Notes.} Same as Table\;\ref{tab:appuppertr}.}
    \end{tabular}
	\normalsize
	\label{tab:applowertr}
\end{table*}

The ratios for the lines around 916\,\AA\ emitted by this ion are not only significantly different within the doublet, but both are far from observations. The high observed intensity of the line at 915.68\,\AA\, relative to predictions, most likely suggests a blend with an unidentified line. However, this cannot account for the observed 916.70\,\AA\ value being much weaker than predictions. Cross redistribution of intensity between the two lines could be a factor to some degree. The position of the lines right at the edge of the Lyman continuum means that subtracting the background could be problematical. There are no other observations to check the values from \citet{parenti2005}. Photo-absorption by Rydberg levels in hydrogen could be another factor affecting the observed intensities. The doublet forms at the same temperature as the \ion{O}{ii} lines around 833\,\AA, and those lines have similar predicted to observed intensity ratios as the 916.70\,\AA\ line.

The models for Si and S in \citet{dufresne2021picrm} show that the neutrals for both elements are completely depleted throughout most of the chromosphere. Emission from the singly-charged ions would require the same modelling as other lines forming in the chromosphere. Even in hydrostatic equilibrium, \citet{lanzafame1994} demonstrates that the line ratios of \ion{Si}{ii} change as a result of opacity when radiative transfer is included. Despite this, some general points can be made based on the present results.

The intensity for the \ion{Si}{ii} 1816.93\,\AA\ inter-combination line in Tab.\;\ref{tab:apptrchromo} is taken from \citet{nicolas1977}. The observation is taken from 300\arcsec\ inside the white light limb. Their \ion{Si}{iii} inter-combination line observations are only enhanced by 1.4 between disc centre and this point. It suggests that the observation at disc centre for the 1816.93\,\AA\ line might not be significantly different, especially because the \ion{Si}{ii} line shows much lower limb brightening. \citet{nicolas1977} finds that the \ion{Si}{ii} doublet at 1526.71\,\AA\ and 1533.44\,\AA\ is optically thick in all parts of the atmosphere. This would explain why the observed intensity ratio is almost 1:1, when in the optically thin limit it would be 1:2, but it could also indicate a blend in the shorter wavelength line. Both lines form slightly higher in wavelength than the \ion{Si}{i} continuum, which could affect their formation.

\begin{figure}
	\centering
	\includegraphics[width=8.4cm]{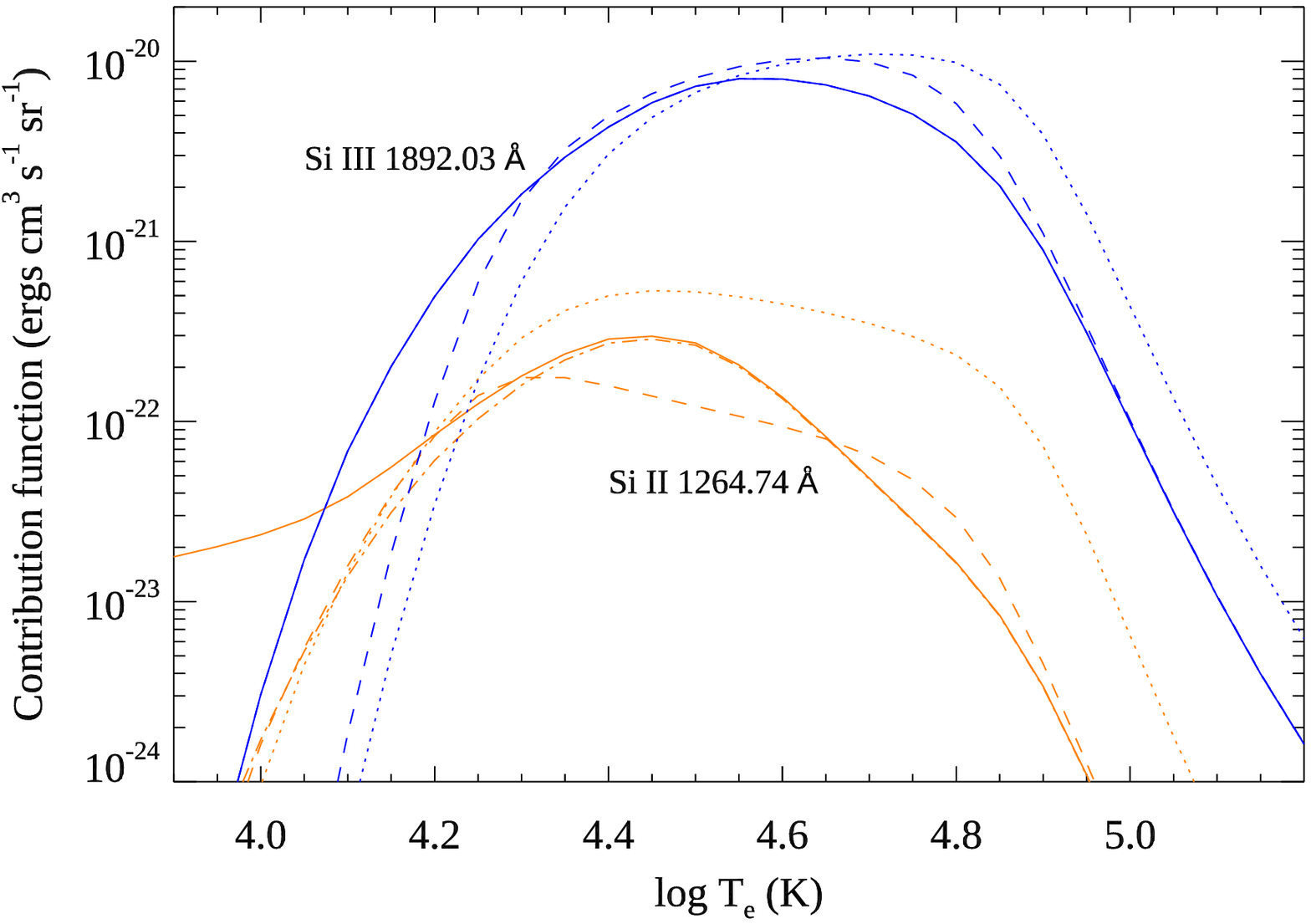}
	\caption[Contribution functions for silicon lines]{Contribution functions of \ion{Si}{ii} 1264.74\,\AA\ (orange) and \ion{Si}{iii} 1892.03\,\AA\ (blue) lines: solid line - full model including PE, dash-dotted - full model without PE, dashed - electron collisional model, and dotted - \textsc{Chianti} v.9.}
	\label{fig:sicontribs}
\end{figure}

\citeauthor{nicolas1977} also find that the multiplet near 1263\,\AA\ is optically thick in all regions of the atmosphere. The 1265.01\,\AA\ line forms deeper in the atmosphere than the other lines in the same multiplet, according to \citeauthor{lanzafame1994}. \citet{dufresne2021picrm} highlights how the upper level which emits the 1264.75\,\AA\ line is enhanced by photo-excitation (PE) more than the upper level emitting the two other lines in the multiplet. The contribution function of this line, shown in Fig.\;\ref{fig:sicontribs}, indicates how potentially the upper level of the line is strongly populated at low temperatures. In all of the models used here, there is no consistency in the predicted to observed ratios for this multiplet. The same also applies to the multiplet around 1194\,\AA.

The one set of \ion{Si}{ii} lines for which the predicted to observed intensity ratios are similar to each other is for that emitted around 1306\,\AA. \citet{lanzafame1994} finds that, although these lines form at similar depths as the 1530\,\AA\ doublet, they emit efficiently in the core of the line in the lower transition region, and so will experience less scattering. \citet{nicolas1977} finds that their profiles only exhibit self-reversal at the limb. It may explain why predictions for these lines from the full model appear to be in reasonable agreement with observations.

The effect of the new ion balances on the formation temperatures of the \ion{S}{ii} lines is also shown in Tab.\;\ref{tab:apptrchromo}. Changes in the ion balance cause the intensities of the resonance lines near 1254\,\AA\ to rise by 70 per cent with the full model, and they now are a factor of two higher than observed. It is interesting to note that their predicted to observed ratios are now the same as those of the \ion{C}{ii} 904.14\,\AA\ and \ion{O}{ii} 833\,\AA\ lines, which form around the same temperature. In this case, however, hydrogen absorption could not cause their observed intensities to be much weaker than predicted.

The predictions for the 1102.30\,\AA\ and the 937.69\,\AA\ lines are close to observations using the coronal approximation, while the latter line drops notably in the full model. The observations for these two lines are from different sources. The latter line is a very weak line, and forms at lower temperature than the \ion{N}{ii} lines around 916\,\AA; it could also be affected by photo-excitation of Rydberg levels in hydrogen. The alternative intensities for all of the \ion{S}{ii} lines are from \citet{parenti2005}, except for the ratio of 2.11 for the 1253.81\,\AA\ line, which is from \citet{vernazza1978}. From these, the wide differences in observations of \citeauthor{parenti2005} can be seen for this ion.

The only observation for \ion{Ne}{ii} is from \citet{andretta2014}. This is very weak and is a blend with \ion{Ti}{xii}, as indicated by its effective temperature. Because, however, the same ion fractions for \ion{Ti}{xii} from \textsc{Chianti} are used in each of the models, the change in predictions is due to the emission from \ion{Ne}{ii}. Considering the predicted intensity from the \ion{Ne}{ii} line alone, compared to the coronal approximation, it decreases by 50 per cent in the electron collisional model and by a factor of two in the full model. The results are further from observations with the present models, but this is a very weak line.

\begin{table*}
	\caption[width=0.7\textwidth]{Comparison of predicted and observed quiet Sun radiances for other transition region-chromosphere boundary lines.}
	\centering
	\begin{tabular}{p{1.3cm}>{\raggedright}p{0.7cm}>{\raggedleft}p{1.3cm}>{\raggedleft}p{1.3cm}>{\centering}p{0.7cm}>{\raggedright}p{0.8cm}>{\centering}p{0.7cm}>{\centering}p{0.7cm}>{\centering}p{0.7cm}>{\centering\arraybackslash}p{1.9cm}}
			\hline\hline \noalign{\smallskip}
			Ion & Seq & $\lambda_{\rm obs}$ & $I_{\rm obs}$ & $T_{\rm c}$ & $T_{\rm f}$ & $R_{\rm c}$ & $R_{\rm e}$ & $R_{\rm f}$ & $R_{\rm f}^{\rm alt}$ \\
			\noalign {\smallskip} \hline \noalign {\smallskip}

\ion{N}{ii} & C & 1083.99 & $11.4^a$ & 4.50 & 4.38 & 0.70 & 0.81 & 0.78 & 0.83 - 1.04 \\
\ion{N}{ii}* & C & $^{sb}$1084.58 & $26.0^a$ & 4.50 & 4.38 & 0.92 & 1.06 & 1.02 & 1.16 - 1.29 \\
\ion{N}{ii} & C & 1085.54 & $11.1^a$ & 4.50 & 4.38 & 0.56 & 0.65 & 0.62 & 0.73 - 0.90 \\
\ion{N}{ii}* & C & 1085.71 & $47.2^a$ & 4.50 & 4.38 & 0.76 & 0.88 & 0.85 & 0.50 - 1.09 \\
\ion{N}{ii} & C & 915.68 & $11.0^d$ & 4.55 & 4.48 & 0.25 & 0.25 & 0.21 &  \\
\ion{N}{ii} & C & $^{sb}$916.70 & $8.0^d$ & 4.58 & 4.52 & 1.85 & 1.81 & 1.53 & \\
\ion{Si}{ii} & Al & 1816.93 & $4365.0^g$ & 4.17 & 3.98 & 0.19 & 0.32 & 0.99 &  \\
\ion{Si}{ii} & Al & 1526.70 & $123.0^b$ & 4.23 & 4.03 & 0.31 & 0.43 & 0.89 & \\
\ion{Si}{ii} & Al & 1533.44 & $126.0^b$ & 4.23 & 4.03 & 0.60 & 0.82 & 1.72 & \\
\ion{Si}{ii} & Al & 1304.38 & $38.8^b$ & 4.27 & 4.08 & 0.41 & 0.49 & 0.77 & 0.71 - 1.25 \\
\ion{Si}{ii} & Al & 1309.32 & $69.7^b$ & 4.27 & 4.08 & 0.40 & 0.47 & 0.75 & 0.67 \\
\ion{Si}{ii} & Al & 1260.42 & $42.9^b$ & 4.29 & 4.11 & 1.26 & 1.38 & 1.92 & 2.59 \\
\ion{Si}{ii} & Al & 1264.75 & $123.0^b$ & 4.29 & 4.11 & 0.78 & 0.85 & 1.19 & 0.89 - 2.10 \\
\ion{Si}{ii} & Al & 1265.01 & $34.1^b$ & 4.29 & 4.11 & 0.29 & 0.32 & 0.45 & 0.29 - 0.66 \\
\ion{Si}{ii} & Al & 1193.31 & $10.7^d$ & 4.30 & 4.14 & 2.55 & 2.62 & 3.29 & \\
\ion{Si}{ii} & Al & 1197.41 & $8.6^d$ & 4.30 & 4.14 & 1.65 & 1.69 & 2.12 & 1.55 - 2.17 \\
\ion{Si}{ii} & Al & 1190.43 & $14.2^d$ & 4.31 & 4.14 & 0.92 & 0.94 & 1.16 & 0.62 - 0.90 \\
\ion{Si}{ii} & Al & 1194.49 & $23.0^d$ & 4.32 & 4.15 & 2.97 & 3.03 & 3.75 &  \\
\ion{Si}{ii} & Al & $^{sb}$992.65 & $1.7^d$ & 4.50 & 4.46 & 3.46 & 2.67 & 2.51 &  \\
\ion{S}{ii} & P & 1253.81 & $14.2^b$ & 4.31 & 4.13 & 1.37 & 1.57 & 2.38 & 2.11 - 2.82 \\
\ion{S}{ii} & P & 1259.52 & $21.2^b$ & 4.31 & 4.13 & 1.25 & 1.44 & 2.18 & 0.59 \\
\ion{S}{ii} & P & 1250.59 & $7.0^b$ & 4.32 & 4.14 & 1.36 & 1.55 & 2.29 & 3.16 \\
\ion{S}{ii} & P & 1102.30 & $4.4^f$ & 4.44 & 4.33 & 0.97 & 0.93 & 0.80 & 4.60 \\
\ion{S}{ii} & P & 937.69 & $1.8^d$ & 4.46 & 4.37 & 1.01 & 0.78 & 0.57 & \\
\ion{Ne}{ii} & F & $^{bl}$460.70 & $1.3^h$ & 5.83 & 5.95 & 0.89 & 0.70 & 0.63 &  \\
			\noalign{\smallskip}\hline
			\noalign{\smallskip}
    \multicolumn{10}{p{0.8\linewidth}}{\textbf{Notes.} Same as Table\;\ref{tab:appuppertr}}
		\end{tabular}
	\normalsize
	\label{tab:apptrchromo}
\end{table*}

\section{Assessment of model parameters}
\label{sec:pressure}

In various diagnostic works on the TR a range of pressures have been determined from density-sensitive line ratios. These typically lie in the range $3\times10^{14} \,-\, 3\times10^{15}$\,cm$^{-3}$\,K. In \citet{warren2005} the ratio of 759.43\,\AA/761.99\,\AA\ line intensities in \ion{O}{v} gives a pressure of $1.3\times10^{15}$\,cm$^{-3}$\,K, while the \citet{wilhelm1998} intensities from the same ratio point to a pressure of $2\times10^{15}$\,cm$^{-3}$\,K. Other lines in \citet{warren2005} forming at higher temperature do not give such a high pressure. However, to check whether higher pressures could affect the results for such anomalous ions as \ion{O}{vi}, for example, the ion balances in the electron collisional models and the DEM were completely re-run using a pressure of $1.0\times10^{15}$\,cm$^{-3}$\,K. (Only the electron collisional models were run in this case because the higher temperature lines are not affected by charge transfer and photo-ionisation.) In most cases the predicted to observed ratios were unchanged at the higher pressure; the biggest change was 2 per cent. This is largely because ion fractions do not change to a large degree between these QS pressures and the DEM will adjust to the new ion populations. Furthermore, many of the high temperature line intensities show changes of only a few per cent when using the electron collisional model compared to the density-independent coronal approximation.

\begin{figure}
	\centering
	\includegraphics[width=8.4cm]{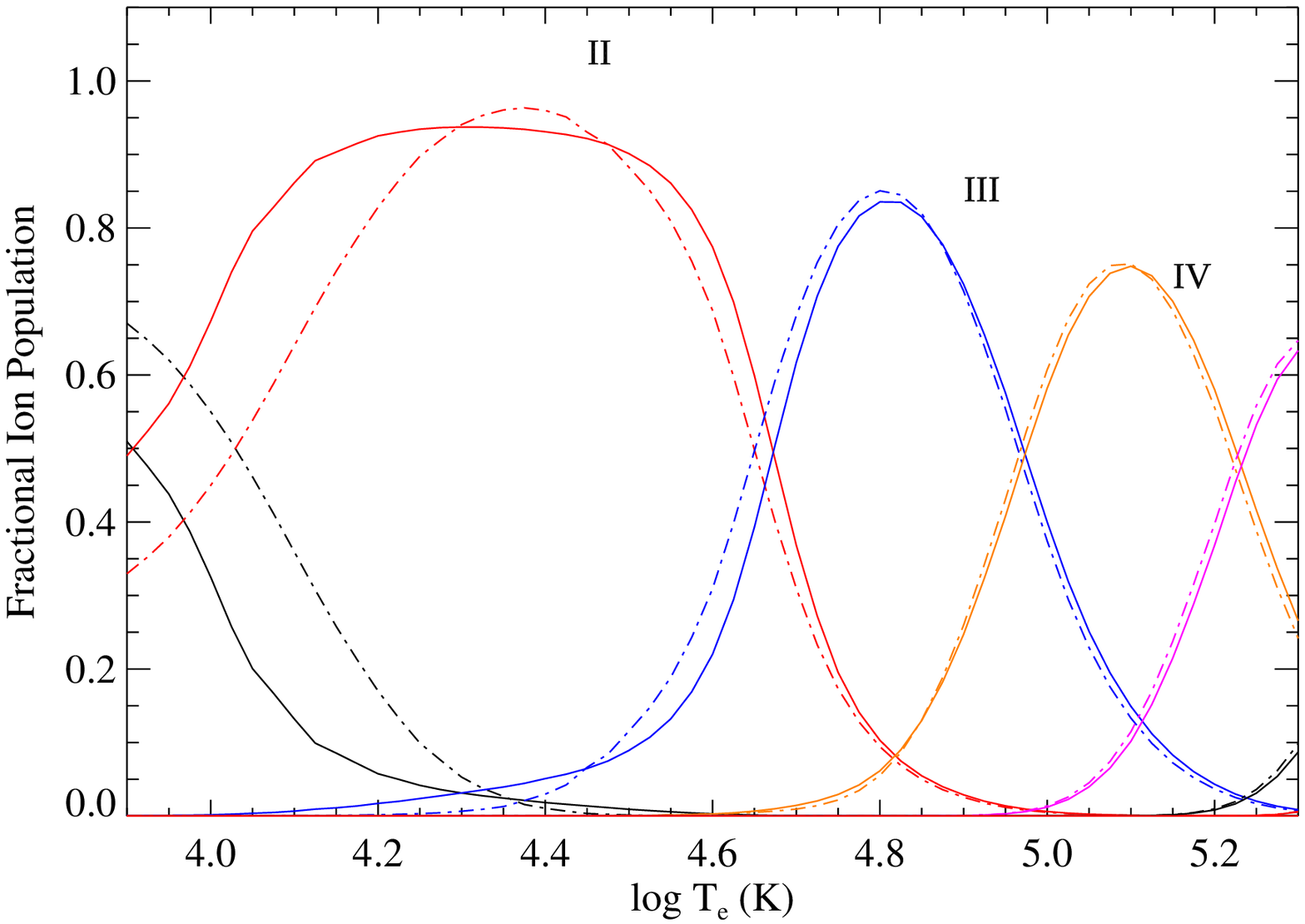}
	\caption[Ionisation equilibrium of oxygen using different model atmospheres and pressures]{Ionisation equilibrium of oxygen using different model atmospheres and pressures with PI and CT: solid line - with PI and CT at pressure $3\times10^{14}$\,cm$^{-3}$\,K and model atmosphere from \citeauthor{avrett2008}, dot-dashed - with PI and CT at pressure $9\times10^{14}$\,cm$^{-3}$\,K and model atmosphere from \citeauthor{fontenla2009}.}
	\label{fig:fontenla}
\end{figure}

Lower in the TR, different pressures could alter the ion balances because photo-induced processes are independent of density, and charge transfer is dependent on both density and hydrogen abundance. The hydrogen abundance also depends on the model atmosphere. To give an indication of how much these factors might change the ion balances and line emission, the full oxygen and silicon models were re-run using the model atmosphere of \citet{fontenla2009}. The data are taken from their model B, which is for a cell centre and exhibits a pressure of approximately $9\times10^{14}$\,cm$^{-3}$\,K over much of the TR. The results for the full model are shown for oxygen in Fig.\;\ref{fig:fontenla} and are compared with the model from \citet{dufresne2021pico}, which uses hydrogen abundances from \citet{avrett2008}.

Overall, the biggest change to the ion balance is in the upper chromosphere, where the charge transfer effect is at its strongest. The effect is similar for silicon, for which the peak abundance of \ion{Si}{iii} increases but the ion forms over a narrower temperature range. The ion fractions using the \citeauthor{fontenla2009} model atmosphere were used to recalculate intensities using the above DEM from the full models. (The \ion{O}{ii} and Si lines were not used to fit the DEM and so the DEM will be unaffected by the changes.) As with all lines in this work, such changes affect lower temperature lines differently than higher temperature lines. Broadly, however, the \ion{O}{ii} and \ion{O}{iii} lines decreased on average by 10 per cent, while the \ion{Si}{iii} lines increased on average by 10 per cent. This is not sufficient to alter the outcome of the present work.

Emission for many low charge ions could be affected if there are areas of lower pressure along the line of sight, such as in prominences. Testing the collisional radiative models at the pressure of $6\times10^{13}$\,cm$^{-3}$\,K, which is given by \citet{parenti2019} as a possible value for prominences, but still using the model atmosphere of \citet{avrett2008}, shows that \ion{O}{ii} would be further depleted by photo-ionisation. Although \ion{O}{iii} is enhanced at low temperature because of this, the peak in abundance of \ion{O}{iii} is also reduced by photo-ionisation. So, the \ion{O}{iii} line intensities may not be adversely affected, while the \ion{O}{ii} line intensities should decrease. \ion{O}{iv} shows an enhancement at lower temperature, highlighting that its inter-combination lines, the strongest of which is under-predicted in the present model, could produce enhanced emission in these conditions. This indicates how lower pressures could alter emission, but more detailed modelling of these conditions would be required.

\section{Assessment of discrepancies found in SUMER observations}
\label{sec:doschek}

The \ion{N}{iii} lines are one case in which \citet{doschek1999} find almost a factor of two discrepancy between observation and theory. It has been highlighted in \citet{dufresne2020} how \citeauthor{doschek1999} normalised the theoretical intensity of each line to the predicted intensity of a reference line, usually an inter-combination line. They also normalise the observations of each line to the observation of the same reference line, and then compare the ratios. For oxygen, the predictions for the inter-combination lines being a factor of two lower than observed caused the factor of two discrepancies \citeauthor{doschek1999} find between the predicted and observed ratios for all of the \ion{O}{iv} and \ion{O}{v} lines. In the case of \ion{N}{iii} they use the 989.82\,\AA\ line to normalise all the other results. If the same method is used here the 764.36\,\AA\ line, which is the line furthest from observations, would show an apparent discrepancy of about 1.6 between observations and predictions when using any of the present atomic models. So, the problems \citeauthor{doschek1999} find for this ion are a consequence of their method of normalising the observations separately from the predictions.

The lines from \ion{S}{iii-v} are those for which \citet{doschek1999} find the most discrepancies in the UV lines they consider. Up to a factor of five difference is found between theory and observations. For \ion{S}{iii} the only lines in common with this work are those at 1194.07\,\AA\ and 680.70\,\AA. In theirs and the present work, both lines show good agreement. The \ion{S}{iv} lines show good consistency here and form at similar temperatures to each other. Consequently, there is not a problem with the line they chose to normalise the others with, nor should their assumption of isothermal conditions affect the results.

The \ion{S}{v} 786.47\,\AA\ line is the line for which \citet{doschek1999} find the greatest discrepancy between theory and  observations. They find a factor of five difference when it is normalised to the inter-combination line. The results here do not show the same discrepancy. It should be considered, though, that prior to the current version of \textsc{Chianti} \citep[v.10,][]{delzanna2021v10}, distorted wave excitation rates from \citet{christensen1986} have been used for this ion. Predictions for the 1199.18\,\AA\ line are a factor of five below observations when using that data. \citet{doschek1999} use the distorted wave data of \citet{pradhan1988}, and so this is likely to explain the factor of five discrepancy \citeauthor{doschek1999} find when normalising the intensities by the inter-combination line. The predictions for the inter-combination line increase by more than a factor of three with the R-Matrix calculations of \citet{menchero2014}, highlighting the impact atomic data can have. All of the electron impact excitation data used here from \textsc{Chianti} for sulphur are from R-Matrix calculations and have been produced since the work of \citet{doschek1999}. This suggests that atomic data was the most likely cause of the discrepancies they found for the sulphur lines.

\bsp	
\label{lastpage}
\end{document}